\documentclass[]{elsarticle}
\usepackage{graphicx,epsf}
\usepackage{hyperref}
\usepackage{xcolor}
\usepackage{psfrag}
\usepackage{amsmath}
\usepackage{amssymb}
\usepackage{psfrag}
\usepackage[english]{babel}
\usepackage[title]{appendix}
\begin{document}
\title{Kinetics of pollutant adsorption onto polymers: the Langevin dynamics study }
\author{Khristine Haydukivska\corref{cor1}}
\ead{wja4eslawa@gmail.com}
\author{Viktoria Blavatska}
\author{Jaroslav Ilnytskyi}
\address{Yukhnovskii Institute for Condensed Matter Physics of the National Academy of  Sciences of Ukraine,1, Svientsitskii Str., 79011 Lviv, Ukraine}
\cortext[cor1]{Corresponding author}

\begin{abstract}
We study adsorption kinetics of low-molecular-weight adsorbate particles on a polymer via the Langevin dynamics. The impact of the polymer molecular weight and of the strength of the polymer-particle attraction are examined. We consider mixtures of $n_p$ chains of the length $N$ with adsorbate, where $n_pN=600$ holds. Analysis includes examination of visualizations and evaluation of adsorption capacity, binding energy, average gyration radius, and asphericity of the polymer-adsorbate complexes. Both adsorption capacity and total binding energy show monotonic, logistic-like behavior; the former reaches a plateau sooner than the latter does, indicating further optimization of the maximum-capacity complex. Binding energy per particle and  asphericity of complexes both demonstrate non-trivial behaviors, with distinct extrema at the intermediate stages of adsorption. The analysis reveals three stages of adsorption: (i) formation of small polymer-particle complexes, (ii) gradual merge of these into a single complex, (iii) stationary state. The stage (i) completes faster for short chains, as in this case fewer conformational rearrangements are required. The stage (ii) completes faster for long chains because of the necklace-like arrangement of small complexes along a long chain, which promotes their merge vs. diffusion-driven merge of the complexes of short chains. Our results indicate that for the rapid removal of pollutants in the high interaction regime, it is more effective to use the set of short polymer chains. On the other hand, long linear chains are optimal for adsorbing pollutants in a weak interaction regime.

\end{abstract}

\maketitle

\section{Introduction}

Cleaning the environment of low-molecular-weight organic pollutants faces much difficulty, as these can migrate easily through water, soil, and biological systems. Well-known examples are various pesticides because of their widespread use in agriculture \cite{Schwarzenbach2003, Wauchope2002, AriasEstvez2008}. One of the most effective ways to decontaminate the environment from them is by adsorption \cite{Nishi2025, AlAqad2025}. The adsorbents of various types can be employed, namely black carbon, clay minerals, and specially engineered adsorbents \cite{Chiou1979, Schwarzenbach2003, Kah2006}, depending on the molecular weight, polarity, hydrogen-bonding ability, hydrophobicity, and other chemical details of a contaminant.

Polymeric adsorbents \cite{RojasFlores2026, Rahmadani2026} offer advantages over conventional mineral and carbon-based materials in the way that both their chemical composition and molecular connectivity are highly tunable. By using such techniques as copolymerization, functionalization, and grafting, and by incorporation of required binding sites, one can achieve the required balance between hydrophobic interactions, hydrogen bonding, electrostatic attraction, and specific molecular recognition \cite{Ulbricht2006, Drioli2010}. High throughput of a polymeric adsorbent is defined not only by the concentration of binding sites but also by their accessibility, which can be affected by chain folding, crosslink density, effects of local aggregation, or solvent shielding. Vice versa, a pollutant may also affect a conformation of linear polymer fragments by intra- and extrachain bridging of the monomers (a pollutant acts as a polymer crowder \cite{Nayar2023}), whereas branching sites may form a natural binding pocket \cite{Kryscio2011}. Therefore, all these factors are interconnected, and, as a result, the problem of predicting an optimal polymer adsorbent structure is non-trivial. In most cases, a polymer adsorber comprises partly immobilized chains that are pregrafted to a surface \cite{Koch2020} and/or crosslinked into a porous matrix of a resin type \cite{Fu2023}. However, high mobility of  polymer-pollutant aggregates of a moderate size, which can be removed by a flow or filtered out, may be beneficial in some applications. Therefore, there is much interest in predicting properties of such aggregates.

The experimental techniques aimed at the characterization of the process of adsorption involve batch-equilibrium experiments, kinetic measurements, column studies, spectroscopic methods, calorimetry, porosity analysis,  etc. \cite{MachadoMachado2026,MouraoCoelho2026}. Equilibrium experiments yield adsorption isotherms providing adsorption capacities, although their fitted parameters do not identify related molecular mechanisms. Neither Freundlich nor Langmuir models \cite{Giles1960,Foo2010} cover the full behavior of flexible polymers, being unable to distinguish the effect of polymer conformations or dynamical creation or destruction of binding sites. Kinetic experiments provide information about the rates of adsorption and desorption. However, the reviews of pesticide sorption indicate that equilibrium coefficients alone are insufficient for predicting transport in heterogeneous environmental systems \cite{Wauchope2002, Kah2006}. Spectroscopic methods are of a great value, as these shed some light on involved interaction types. Infrared and Raman spectroscopy reveal changes undergone by carbonyl, hydroxyl, amine, aromatic, and other functional groups. Molecular mobility and local chemical environment can be traced down by nuclear magnetic resonance, whereas isothermal titration calorimetry can provide binding enthalpies and association constants. Nevertheless, these experimental techniques provide averaged information and cannot distinguish the effects of pollutant diffusion, conformational rearrangement of polymer adsorbent, formation of complexes, etc. at molecular resolution.

Computer simulations, involving real particle dynamics, aim at linking experimentally measured adsorption properties and dynamics with certain mole\-cu\-lar-scale mechanisms. Density functional theory, operating on a quantum level, can be employed to examine geometries and interaction energies of pollutant-monomer complexes as well as check affinity of different functional groups \cite{Mayes1997, Haupt2000, Pichon2007}. Their principal limitation is to cover polymer adsorbent at realistic length and time scales. On the other side of the spectrum are coarse-grained approaches, such as the Brownian and dissipative particle dynamics, that allow one to reach the length and time scales required for describing diffusion, association, dissociation, and transport of both adsorbent and a pollutant. Coarse-graining \cite{MllerPlathe2002, Voth2009, Noid2013} allows simplification of complex interactions by emphasizing the most relevant ones, but requires consistent parametrization approaches.

Molecular dynamics (MD), and the related approach of Langevin dynamics,  can be seen as a reasonable middle ground between the quantum and coarse-grained levels of description, as it both retains relevant atomic details and allows for a reasonably long time description of system dynamics. One may distinguish two types of application of this approach, depending on the focus of a study. Namely, when attempting to model particular chemical compounds, both for polymer adsorbent and a pollutant, the atomistic force fields (e.g., AMBER, CHARMM, OPLS, GROMOS, etc.) and explicit water models can be employed \cite{Brooks2009, Jorgensen1983, Pronk2013}. However, when the main purpose is to establish the roles played by the details of the polymer molecular architecture and of the polymer-pollutant binding strength, then the generic models, typically operating in reduced units, can be used. One can simplify the description further by abandoning fine water structure and mimic it as a viscous thermostatted environment, resulting in the Langevin dynamics.

Such particle-resolved MD simulations are able to establish a wide range of the structure properties, such as pollutant-monomer radial distribution function and coordination number, diffusion coefficients, radius of gyration, asphericity, and relative shape anisotropy of polymer-pollutant aggregate, etc. One also may clearly distinguish whether a pollutant resides predominantly on the external surface of an aggregate or is partially or fully enclosed within it. Moreover, MD is capable of tracing the role played in the formation of such an equilibrium structure by an interplay between a set of relevant factors. One of these is polymer molecular weight, which affects the adsorption by several ways. With its increase, the number of available binding sites also increases, but the polymer may be more prone to collapse, which will render internal binding sites inaccessible, thus reducing adsorbent capacity per molecular mass. Another factor is the increase of the chains' relaxation times, that are proportional to the square of their length, according to the Rouse dynamics \cite{Rouse1953}. This may affect transport dynamics of aggregates and reduce the throughput of polymer adsorbent.

Other factors to mention are chain flexibility, which allows conformational changes towards the formation of more stable multi-valent monomer-pollutant complexes; polymer swelling, collapse, or bridging caused by binding of a pollutant, etc. One should note that the total affinity of a polymer is not a mere sum of its monomers' affinities, as a chain could form a shared binding pocket involving adjacent monomers, produce intrachain and interchain association, or prevent binding between nearby monomers because of steric repulsion \cite{Mosbach1996, Sellergren2001, Pichon2007}. The chemical basis for adsorption is the monomer-pollutant interaction, which may include hydrophobic association, electrostatic attraction, hydrogen bonding, aromatic stacking, etc. The estimates of interaction strengths can be obtained from the quantum-chemical calculations or by using \textit{ad hoc} interaction potentials.

Formation of the polymer-pollutant aggregate is a dynamic process involving translational diffusion, molecular encounter, rearrangement of a solvent, and conformational reorganization of a polymer.  The process involving the reversible binding of low-molecular-weight particles (pollutants or crowders \cite{Nayar2023}) on the polymers leads to crucial conformational reconstruction of macromolecules. In the regime of a good solvent, the neutral flexible polymer chain usually attains a swollen coil conformation, beneficial from the point of view of conformational entropy. Presence of adsorbing particles in solvent, under the condition of strong enough particle-monomer interaction, impacts the conformational equilibrium state of polymer: a transition to globule-like conformations of resulting agglomerates is observed \cite{Garg2025, Huang2021}.

It may start from formation of short-lived univalent monomer-pollutant contacts and then reach a more stable configuration. The binding/unbinding rates are affected by both pollutant and polymer dynamics and the final binding energy. During this process, polymer porosity plays an essential role, which, in the case of flexible polymer adsorbent, is dynamic, in contrast to the static porosity of their rigid inorganic counterparts. Transient pores and channels are formed and disappear as the chain conformations change. A pollutant may be adsorbed within an existing cavity or cause the formation of a new one via reorganizing of adjacent monomers. The shape of the polymer-pollutant aggregate can be characterized by its radius of gyration, asphericity, relative shape anisotropy, and polymer-pollutant contact geometry \cite{Blavatska2011}. These characteristics distinguish between compact spherical, prolate, or oblate globular complexes and rather ``hairy" objects, the factors affecting aggregate stability and its diffusion as a whole.


As follows from a brief overview above, computer simulations can be employed for predicting the dynamics of the polymer-pollutant aggregation, as well as for  examining the internal structure and adsorption capacity of such aggregates. The use of generic particle models allows one to emphasize the universal relations between aggregate size, shape, and interaction strength on one side and adsorption characteristics on another, over the particular chemical details. This generality allows a certain level of transferability across a broad class of polymer-pollutant pairs rather than validity for some specific chemical combination. Contrary to the Monte Carlo methods, the MD allows to trace kinetic pathways related to diffusion, adsorption/desorption rates, and conformation changes directly from the time trajectories. Contrary to the experimental works, where the behavior of such a system is the result of mixing several factors that may be difficult to separate, the polymer structure can be controlled with high precision in terms of molecular weight, connectivity, and solution purity. Such predictive screening may help to identify optimal physical properties for new polymer adsorbents before their chemical synthesis.

The aim of the present work is to model the dynamic processes of adsorption in its most generic terms possible in order to provide an outline for more detailed and specific studies in the future. We focus on studying reversible processes when adsorption is due to physical attraction only. The outline of the work is as follows: first, we introduce the model and its parameters for the molecular dynamics simulations, then we follow with a discussion of the results, and finally, we provide concluding remarks.

\section{Langevin Dynamics Simulations}

We consider a set of cases for the polymer-pollutant mixtures in implicit solvent. Each case is characterized by a certain number of linear polymer chains, $n_p$, each with the degree of polymerization $N$, where the total number of beads in a mixture is constant in all cases and equal to $n_pN=600$. The number of the single bead particles initially in bulk is $n_a(0)=1024$.  Hereafter,  we use the term ``particle'' to describe the pollutant to reflect the generic nature of the model. The index $a$ is used in what follows to denote quantities related to pollutant or adsorbate. Generic bead-spring polymer model \cite{grest1987} is used in the three-dimensional continuous space. All beads, both of polymer chains and of  particles, are of identical mass, $m$, and diameter, $\sigma$. Polymer-polymer and particle-particle (adsorbate-adsorbate) pairs, denoted hereafter as $pp$ and $aa$, respectively, interact via the shifted and truncated pair potential of Weeks-Chandler-Anderson (WCA)
\begin{equation}
 V_{pp,aa}^{\rm WCA}(r) = 4\epsilon
 \left\{
 \begin{array}{ll}
\displaystyle \left(\frac{\sigma}{r}\right)^{12}\!\!\!\! -  \left(\frac{\sigma}{r}\right)^{6}\!\! + 1/4, & r\leq 2^{1/6}\sigma,\\
0,& r>2^{1/6}\sigma,\\
\end{array} 
\right.
\label{wca}
\end{equation} 
here $r$ is a distance between the centers of spherical beads, while $\epsilon$ and $\sigma$ are chosen as the units of energy and length, respectively. Bonding within each chain is represented by the finitely extensible nonlinear elastic (FENE) potential \cite{Kremer_1990}
\begin{equation}
 V_{pp}^{\rm FENE}(r)=- 0.5kr_0^2\ln{[1-(r/{r_0})^2]}.
\label{fene}
\end{equation}
with the bending energy constant $k=30\epsilon/\sigma^2$ and reference bond length $r_0=1.5\sigma$. 

Polymer and particle bead pairs interact via the shifted attractive potential of the similar form to Eq.~(\ref{wca}) but truncated at longer distance of $r_{\rm cut}=2.5\sigma$  
\begin{equation}
 V_{ap}^{\rm LJ}(r) = 4\epsilon x_{ap}
 \left\{
 \begin{array}{ll}
\displaystyle 
\left(\frac{\sigma}{r}\right)^{12}\!\!\!\! -  \left(\frac{\sigma}{r}\right)^{6}\!\! -
\left(\!\frac{\sigma}{r_{\rm cut}}\!\right)^{12}\!\!\!\! + \left(\!\frac{\sigma}{r_{\rm cut}}\!\right)^{6}, & r\leq r_{\rm cut},\\
0,& r>r_{\rm cut},\\
\end{array} 
\right.
\label{ap}
\end{equation} 
Here, the dimensionless parameter $x_{ap}$ denotes the well depth in units of $\epsilon$, it ranges from 2 to 4 in this study.

Numerical simulations are performed via the Langevin dynamics (LD) using the Large-scale Atomic/Molecular Massively Parallel Simulator (LAMMPS) package \cite{lammps}. In this work the velocity Verlet integrator is used  in an $NVT$ ensemble with the temperature maintained by the dumping term with the coefficient $\zeta=0.5\,m\tau^{-1}$, where $\tau = \sqrt{m\sigma^2/\epsilon}$ is the damping time. The integration timestep, $\Delta t = 0.005\tau$, is chosen as a time unit. A cubic simulation box with the dimensions of $L_x=L_y=Lz=85\sigma$ is used with the periodic boundary conditions. Initial state of polymers is generated as a set of  self-avoiding walks (SAW) on a simple cubic lattice using a pivot algorithm, to achieve much faster initial relaxation than that by using the LD \cite{Madras88}. Typical simulation times were $5\cdot 10^7$ LD timesteps.

\begin{figure}[b!]
\centering
\includegraphics[trim=340pt 0pt 350pt 80pt, clip, width=0.42\textwidth]{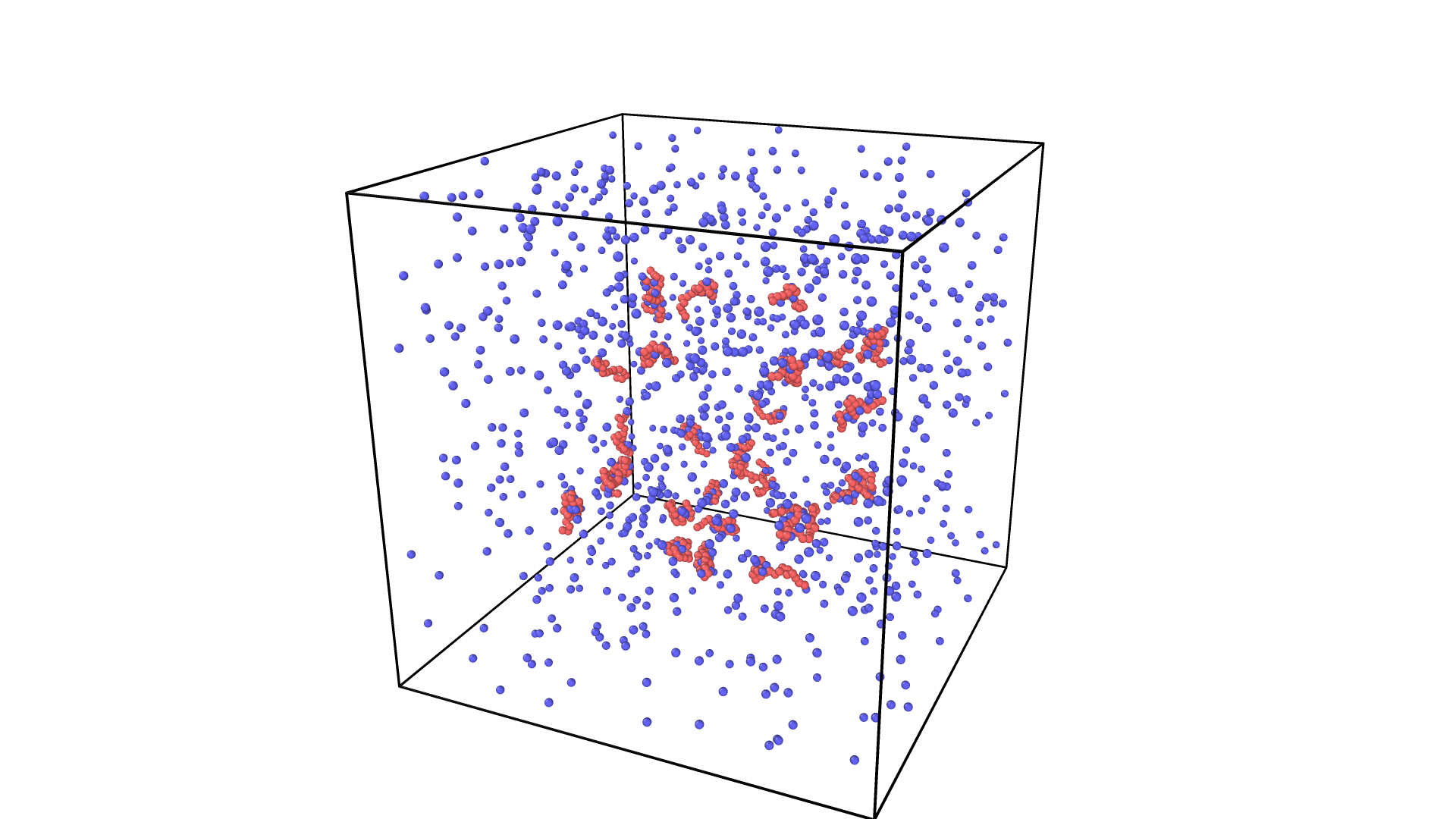}
\includegraphics[trim=340pt 30pt 350pt 80pt, clip, width=0.47\textwidth]{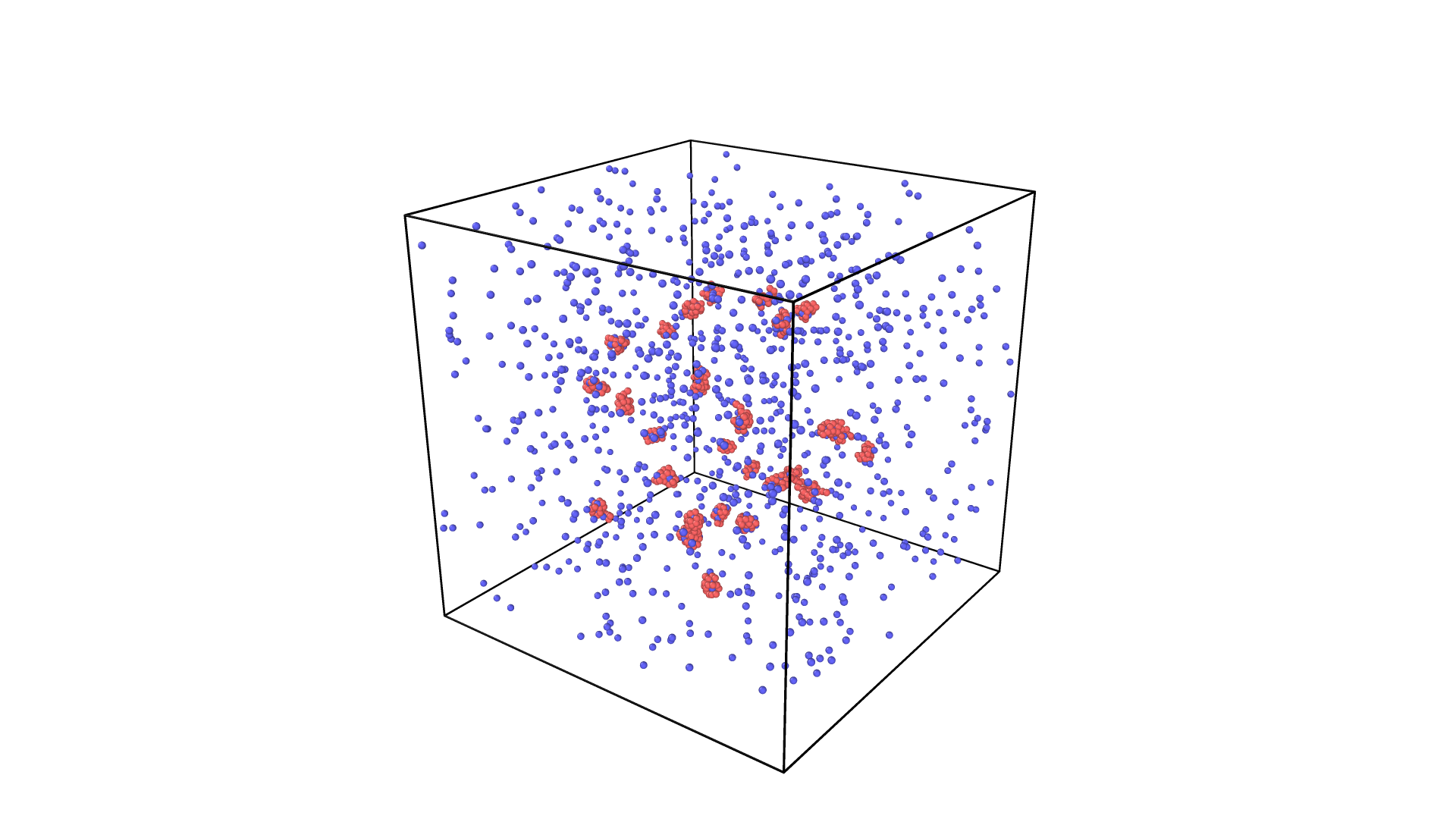}\\
(a)\hspace{12em}(b)\\
\includegraphics[trim=340pt 30pt 350pt 80pt, clip, width=0.45\textwidth]{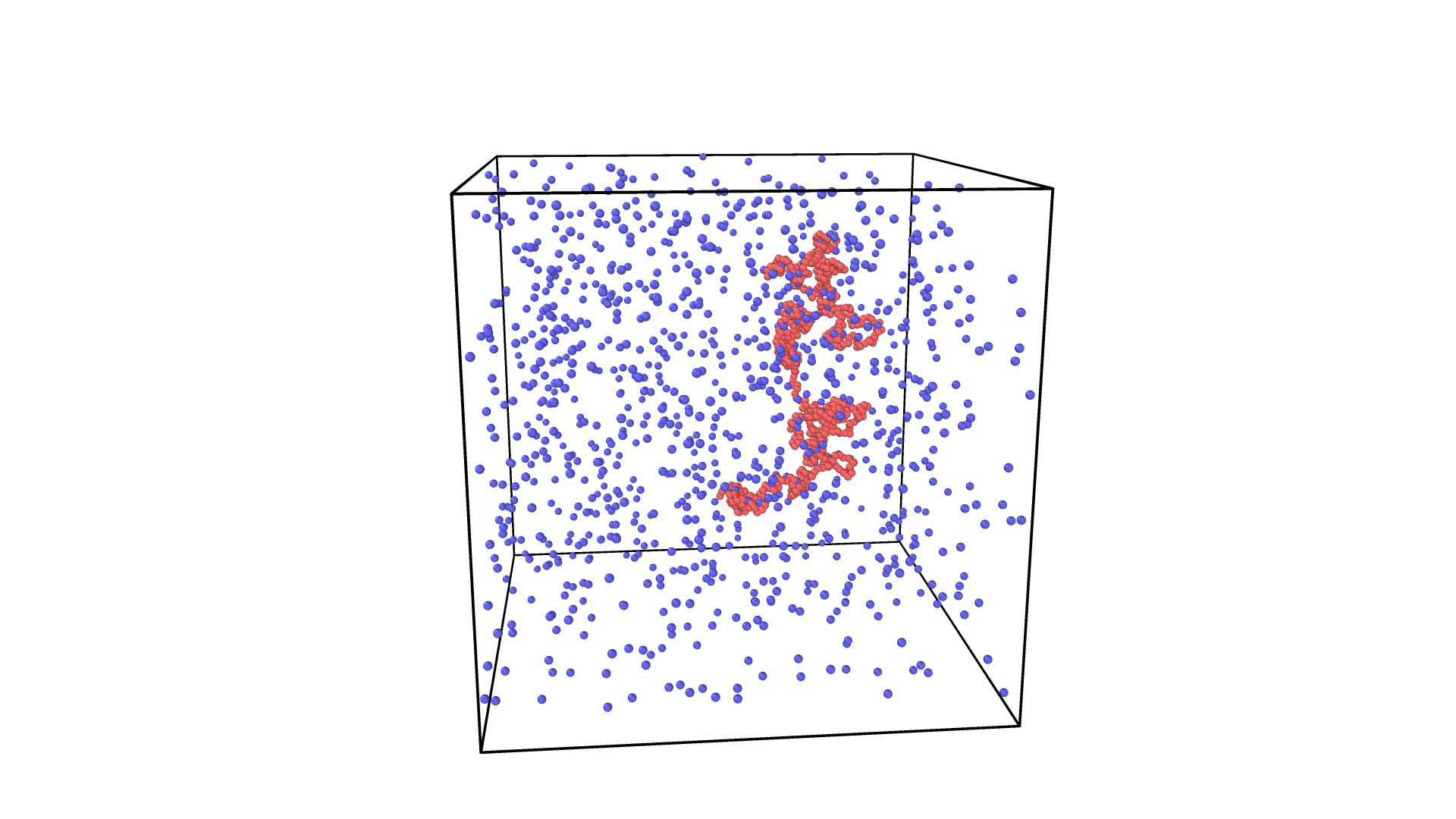}
\includegraphics[trim=340pt 30pt 350pt 80pt, clip, width=0.45\textwidth]{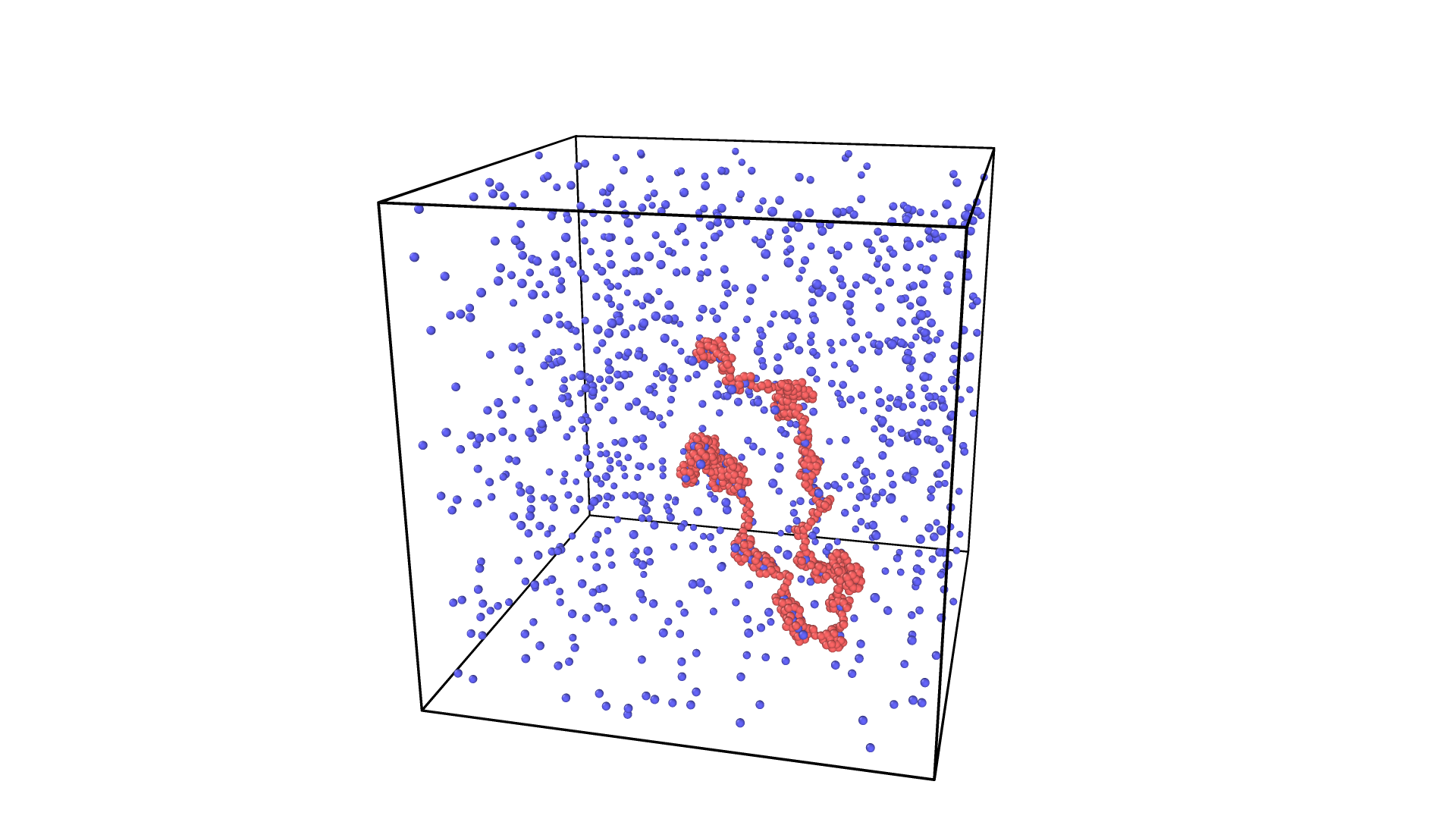}\\
(c)\hspace{12em}(d)\\
\caption{Snapshot of the polymer (shown in red) and particles (shown in blue) at the timestep $1.5\cdot 10^4$. (a) $n_p=30$ short chains of $N=20$ beads at $x_{ap}=2$, (b) the same at $x_{ap}=4$. (c) a single  chain of $N=600$ beads at $x_{ap}=2$, (d) the same as the latter at $x_{ap}=4$.} 
\label{snap1}
\end{figure}

\section{Results and Discussions}

\begin{figure}[t!]
\centering
\includegraphics[trim=340pt 30pt 350pt 80pt, clip, width=0.40\textwidth]{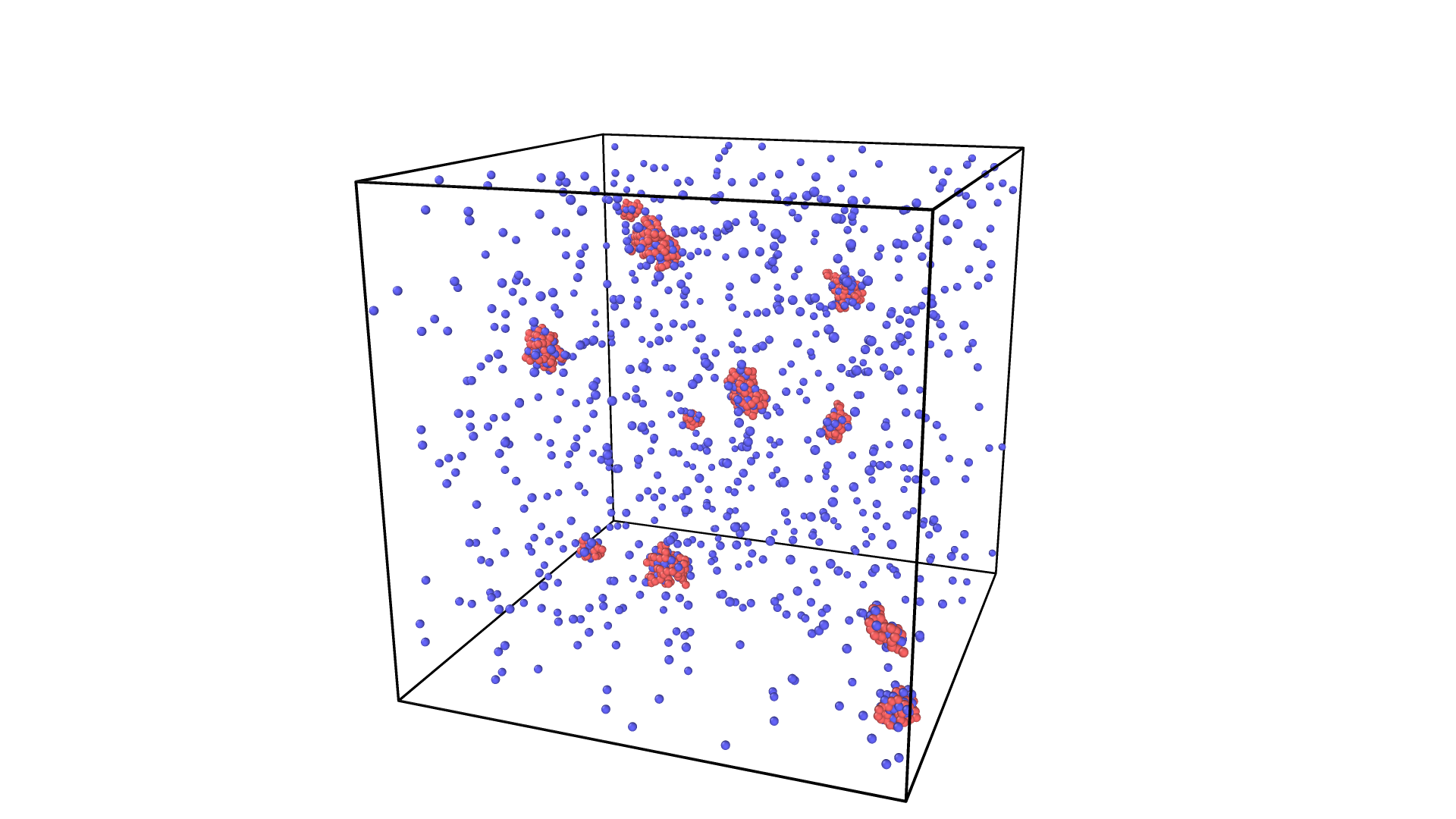}
\includegraphics[trim=340pt 30pt 350pt 80pt, clip, width=0.43\textwidth]{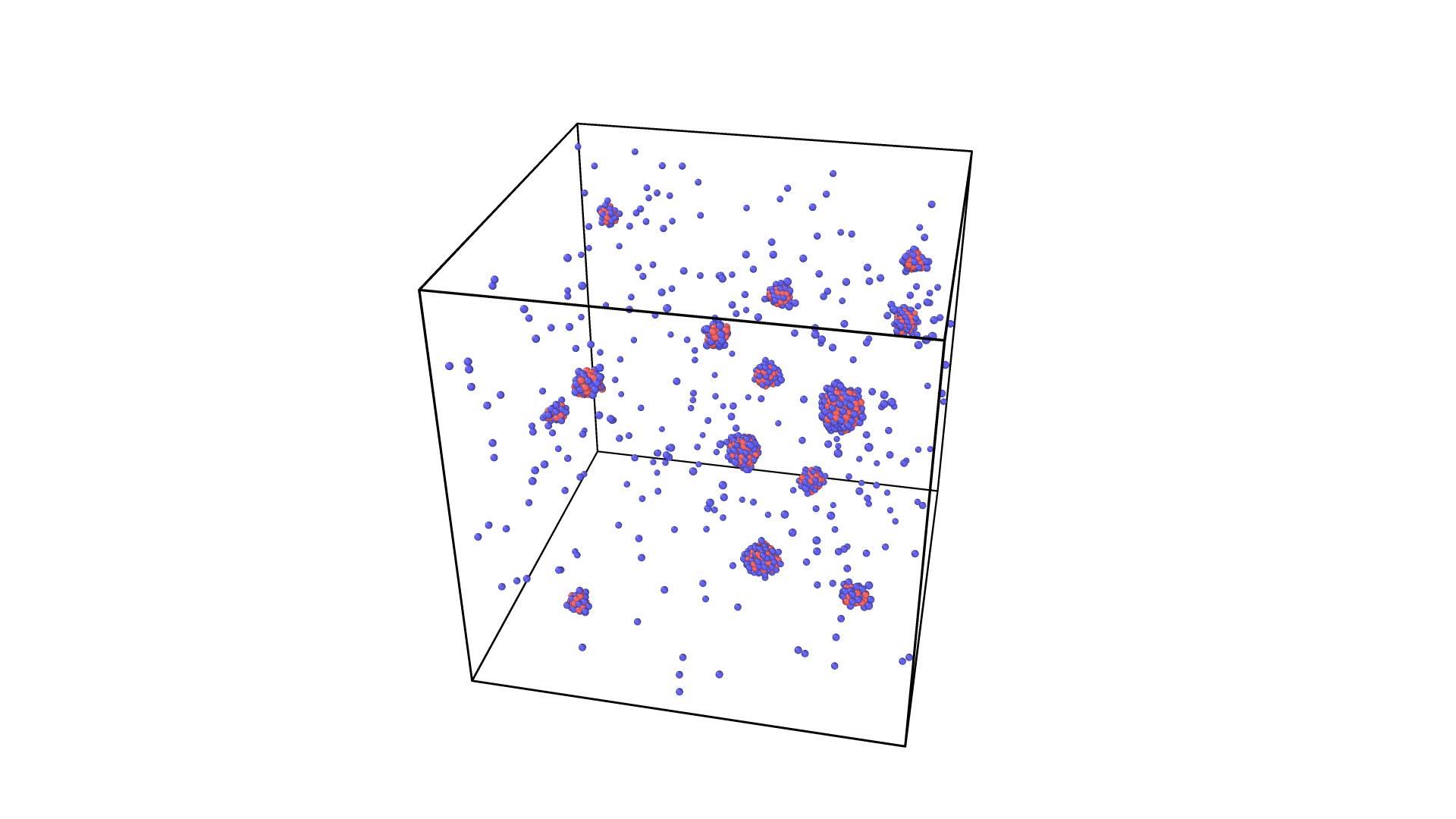}\\
(a)\hspace{12em}(b)\\
\includegraphics[trim=340pt 30pt 350pt 80pt, clip, width=0.40\textwidth]{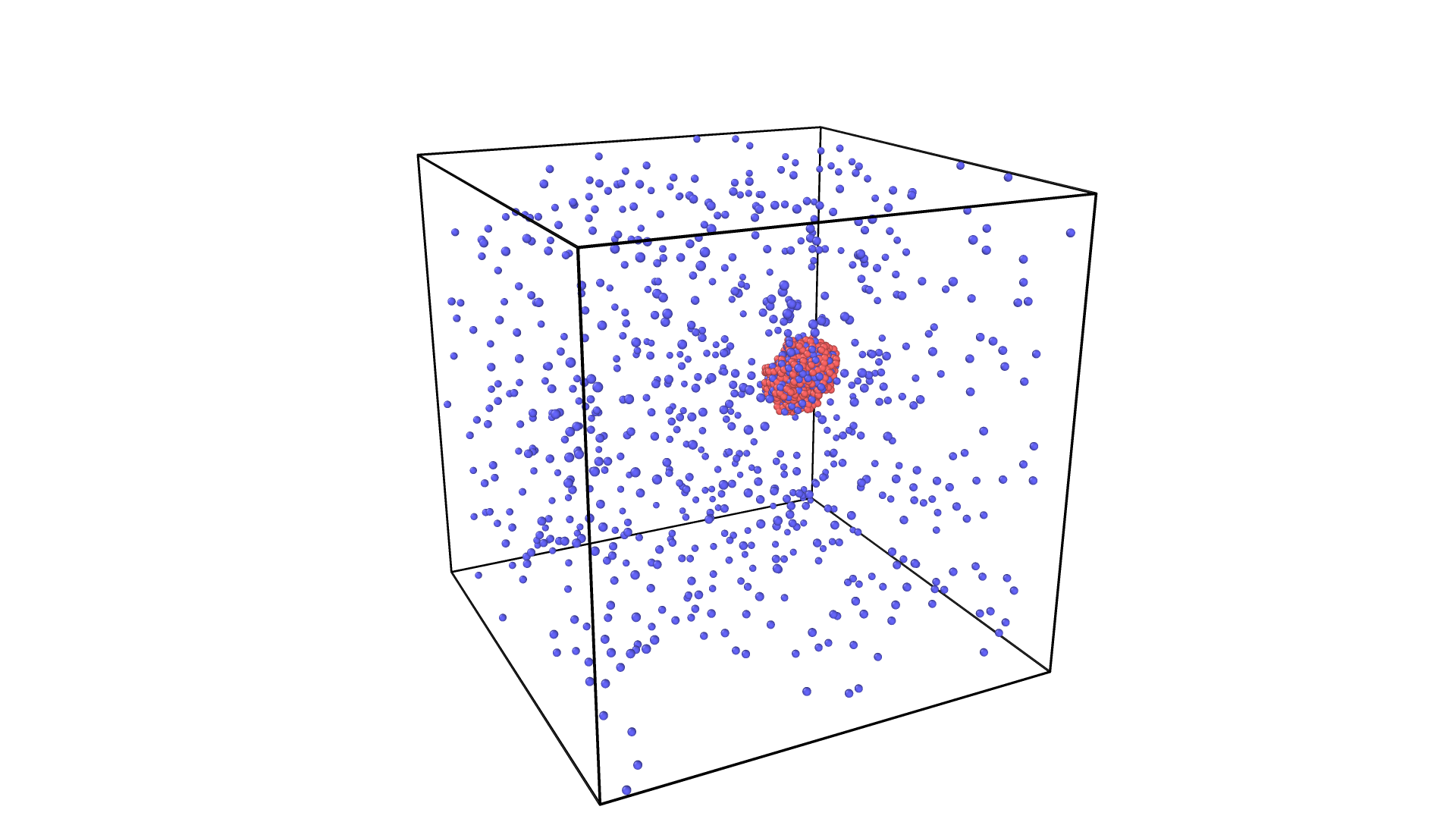}
\includegraphics[trim=340pt 30pt 350pt 80pt, clip, width=0.40\textwidth]{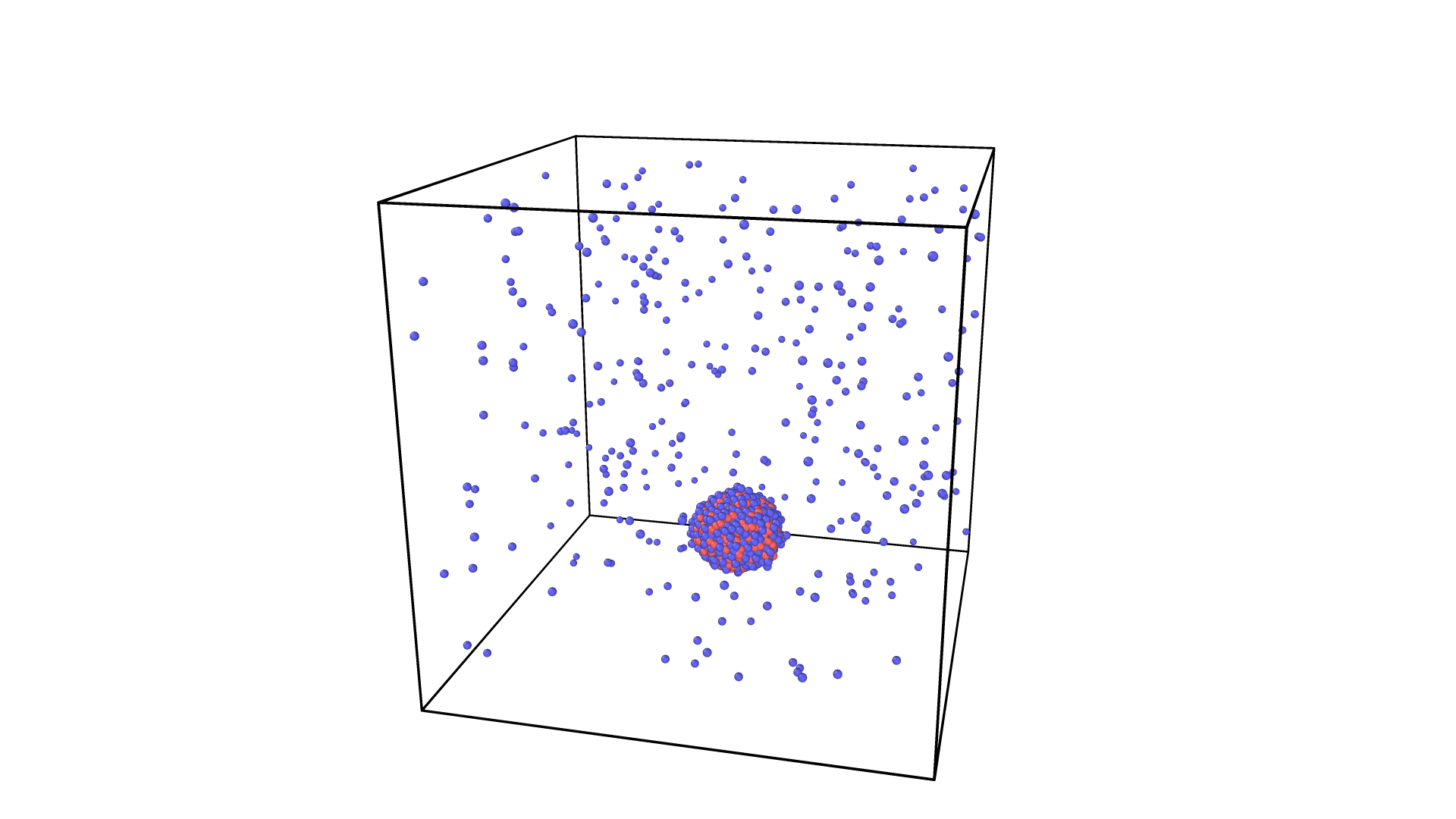}\\
(c)\hspace{12em}(d)\\
\caption{The same as in Fig.~\ref{snap1} but at the timestep of $5\cdot 10^6$.} 
\label{snap2}
\end{figure} 

We start from the analysis of two limiting cases: (i) a set of $n_p=30$ low-molecular-weight (short) chains, each comprising $N=20$ beads, and (ii) a single high-molecular-weight (long) chain of $N=600$ beads. Two values of the well depths of $x_{ap}=2$ and $x_{ap}=4$ are considered.  At an early adsorption stage, $t<10^4$, both systems evolve in time almost identically and independently of $x_{ap}$. In this case only a few particles have been adsorbed on each chain, and the associated energy gain is insufficient to induce some detectable conformational changes.

At the later time step of $1.5\cdot 10^4$ (see Fig. \ref{snap1}), we observe a gradual, ad\-sorp\-tion-induced, folding of chains. Each short chain folds independently of others, and the increase of $x_{ap}$ from $2$ to $4$ promotes more convexity of their shape; see frames (a) and (b). A single long chain, see frames (c) and (d), forms the necklace-like conformation with localized {``nuclei''} (induced by particle-polymer bridging) arranged along the chain, and there is no detectable difference between the cases of $x_{ap}=2$ and $4$, at least for the snapshots being shown.

\begin{figure}[b!]
    \centering
\includegraphics[trim=340pt 30pt 350pt 80pt, clip, width=0.45\textwidth]{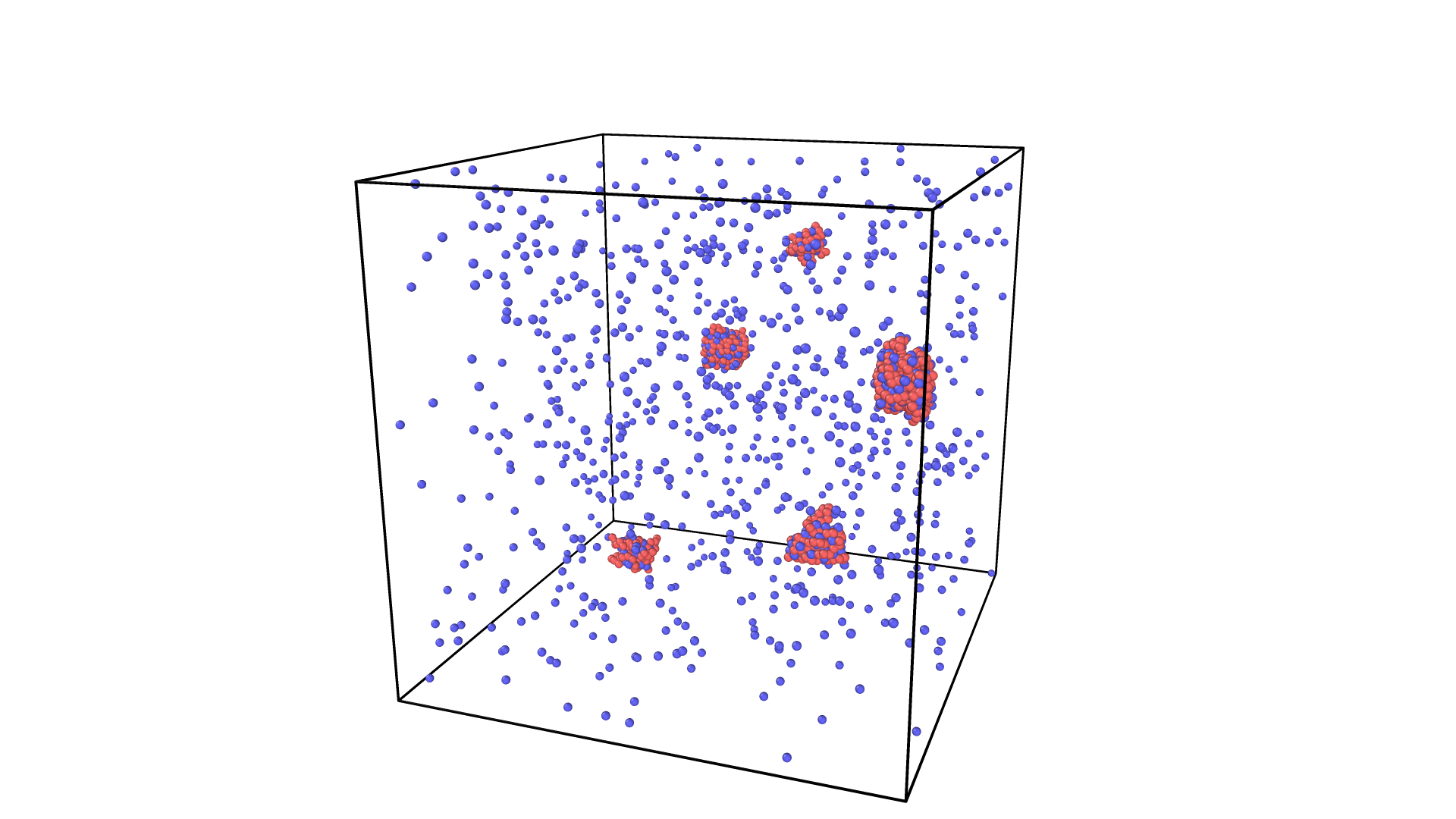}
\includegraphics[trim=340pt 30pt 350pt 80pt, clip, width=0.45\textwidth]{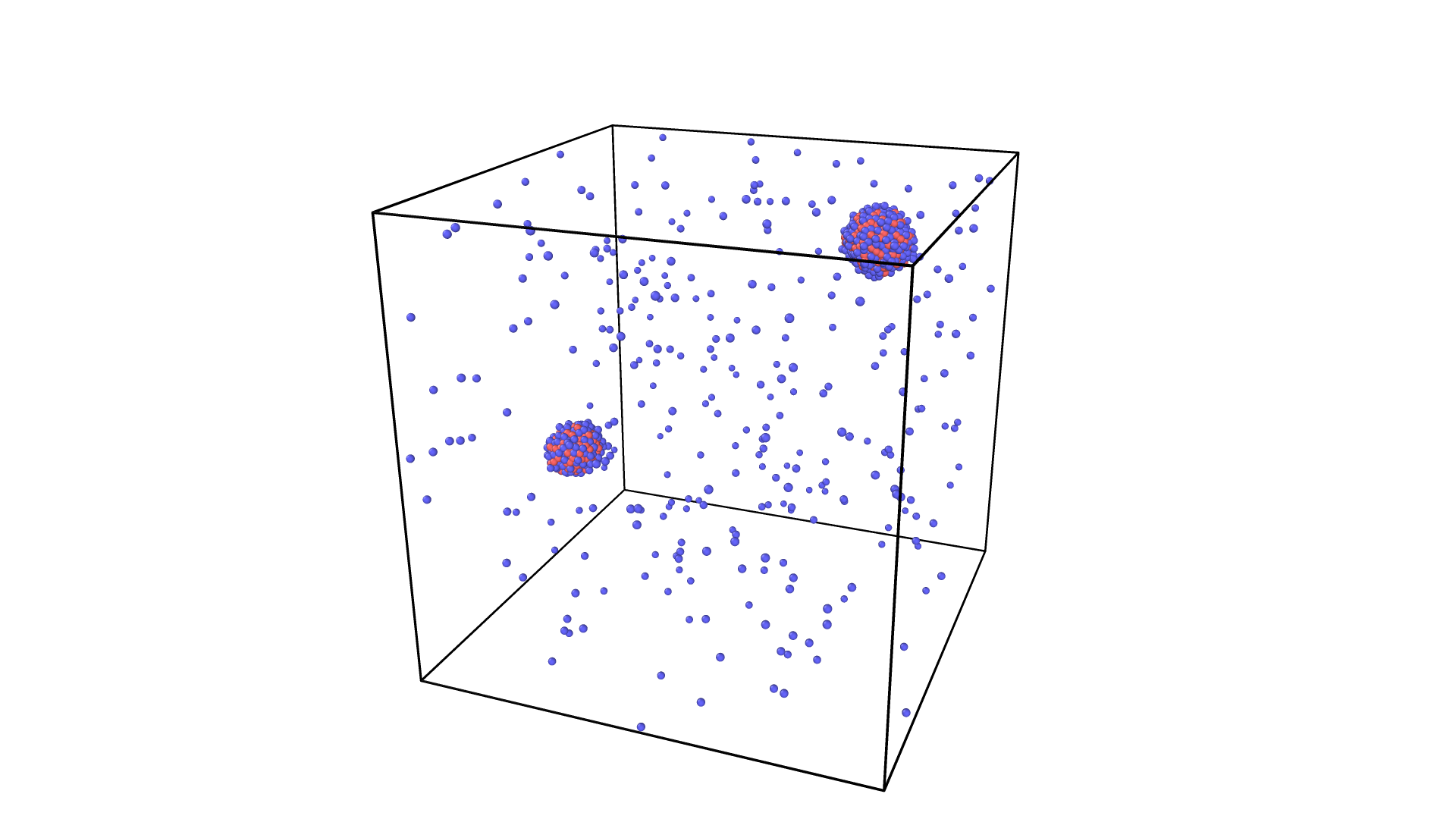}\\
(a)\hspace{12em}(b)\\
\includegraphics[trim=340pt 30pt 350pt 80pt, clip, width=0.45\textwidth]{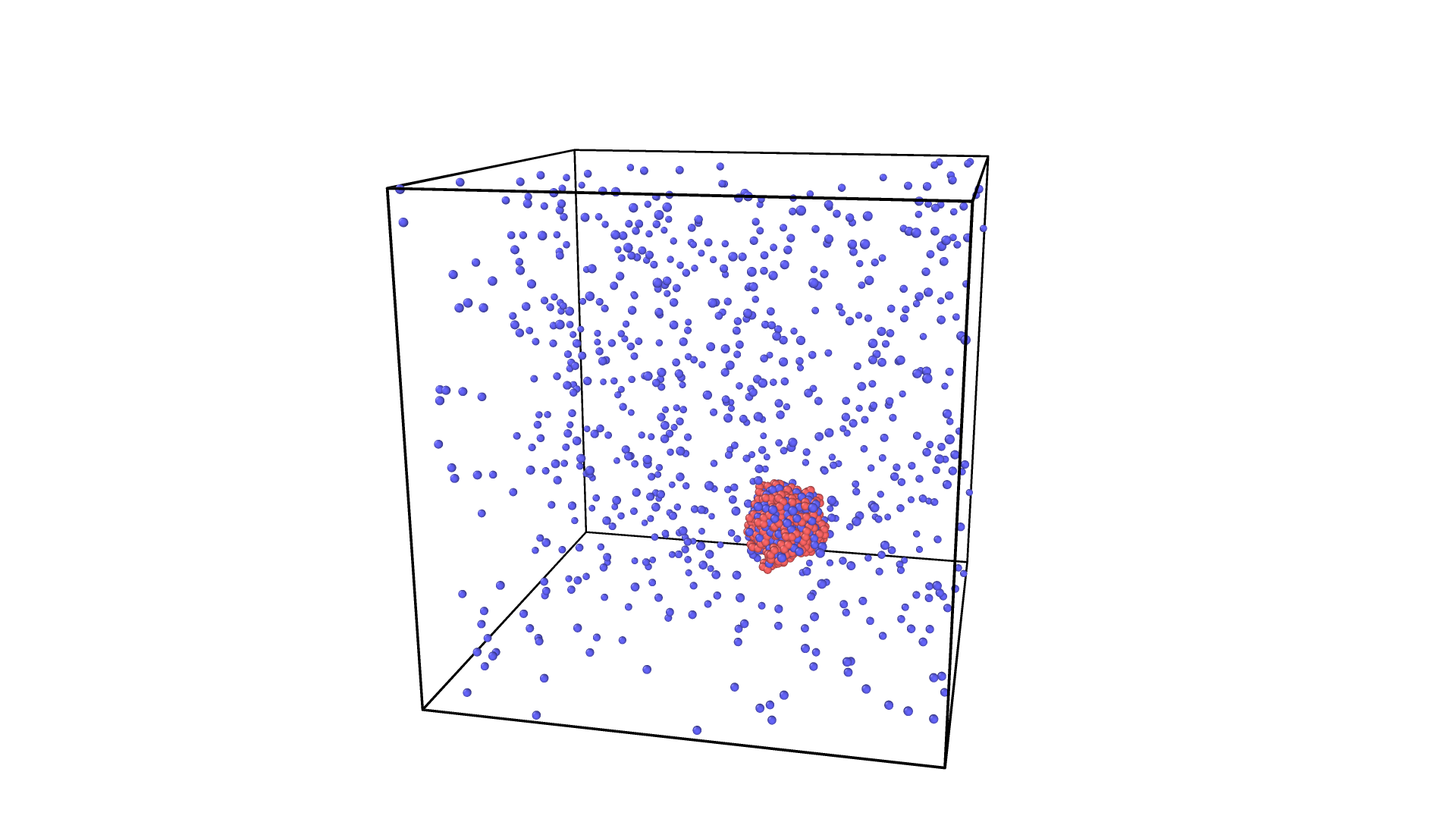}
\includegraphics[trim=340pt 30pt 350pt 80pt, clip, width=0.45\textwidth]{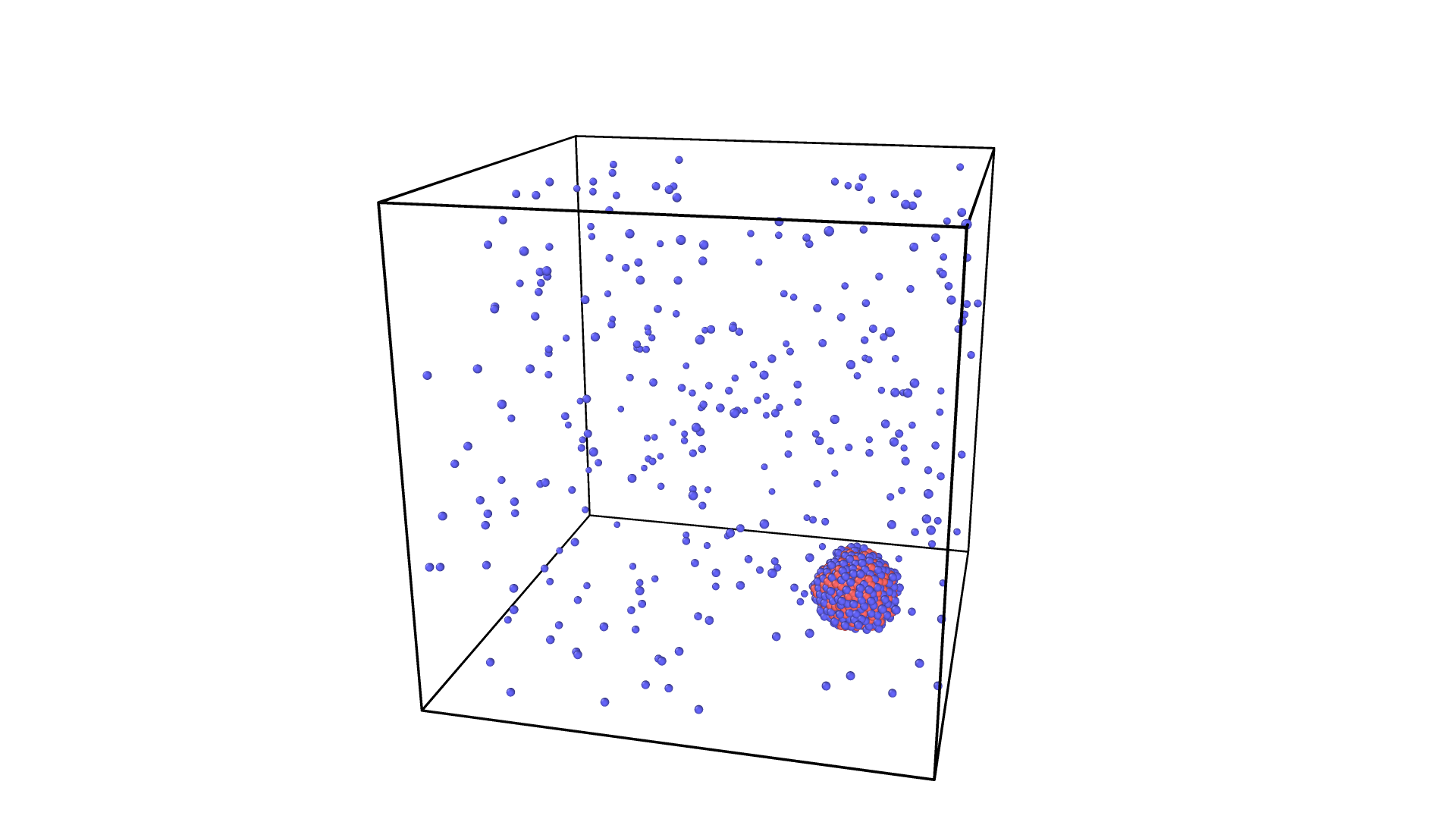}\\
(c)\hspace{12em}(d)\\
\caption{The same as in Fig.~\ref{snap1} but at the timestep of $5\cdot 10^7$.} 
\label{snap3}
\end{figure} 

At the yet later adsorption stage, $t=5\cdot 10^6$, displayed in Fig.~\ref{snap2}, folded short chains are gradually merging into larger complexes; see frames (a) and (b). Similarly to the respective frames of Fig.~\ref{snap1}, the case of $x_{ap}=4$ is characterized by a more convex and more spherical shape. Meanwhile, a single chain already collapses into a single convex agglomerate for both cases of $x_{ap}=2$ and $4$, see frames (c) and (d). For both short and long chains, the complexes are visually more spherical for the case of $x_{ap}=4$, and their surface is covered by adsorbed particles. Those surface particles are very likely to form a bond with only one polymer bead and are held in place due to the strong attractive interaction. { On the other hand, at $x_{ap}=2$ we observe only a few particles on the surface, leading to the hypothesis that a single binding is not enough to hold the particle in long-term. The difference between the times required for a merge of a set of complexes formed of short chains in bulk vs. the nuclei of subchains formed along a single long chain stems from their interconnectivity. In the former case, the process is entirely diffusion driven and thus slow, whereas in the latter case, the complexes are necklaced on a single chain, and this speeds up their merge. }

The typical  near-stationary state of the adsorption process  is presented in Fig.~\ref{snap3}. At this stage there are still five complexes of short chains at $x_{ap}=2$ and two at $x_{ap}=4$, see frames (a) and (b), respectively. Therefore, the merging stage for short chains is approaching the one for a long chain, i.e., a single agglomerate, which already formed at the previous stage as shown in Fig.~\ref{snap2}. Similarly to the latter case, the agglomerate at $x_{ap}=4$ is more spherical than that at $x_{ap}=2$, indicating that the adsorption process may still slowly continue through minor conformational optimization of the agglomerate via slow diffusion of particles from the surface into the core of it. Based upon this visual inspection, one may conclude that the formation of the complexes and their consequent merging are slowed down by both splitting the polymer chain into a set of smaller chains and by reducing the well depth of the particle-polymer attraction potential.

We will proceed now to the quantitative description of the adsorption of particles by polymer chains. One of the most widely used characteristics is the adsorption capacity $q(t)$
\begin{equation}
q(t) = \frac{c_a(0)-c_a(t)}{M}V = \frac{n_a(0)-n_a(t)}{n_pN}.
\end{equation}
Here $c_a(t)$ is a concentration of pollutant particles at time $t$, $V$ is the volume of the simulation box, and $M$ is a total mass of polymer adsorbent. As all the bead types in our model have the same masses, we reduced the standard expression to the fraction between bead numbers. Due to the focus of our study on adsorption dynamics, we will consider the time evolution of $q(t)$, whereas in the experimental studies the limit $q(\infty)$ is typically measured. 

When considered at the level of a lattice polymer model, a set of short chains displays a higher value of $q(\infty)$ compared to the case of a single chain with the same molecular weight \cite{Blavatska2024b}. This effect was explained by the role of the terminal monomers of each chain, which in $3D$ are able to adsorb up to five particles each, whereas the internal monomers can adsorb up to four particles. It is interesting to check whether or not this effect is also present in the off-lattice simulations. On top of that, the LD allows direct study of the adsorption dynamics, whereas the Monte Carlo approach, used in Ref.~\cite{Blavatska2024b}, focused on the stationary state only.

\begin{figure}[b!]
\centering
\includegraphics[width=60mm]{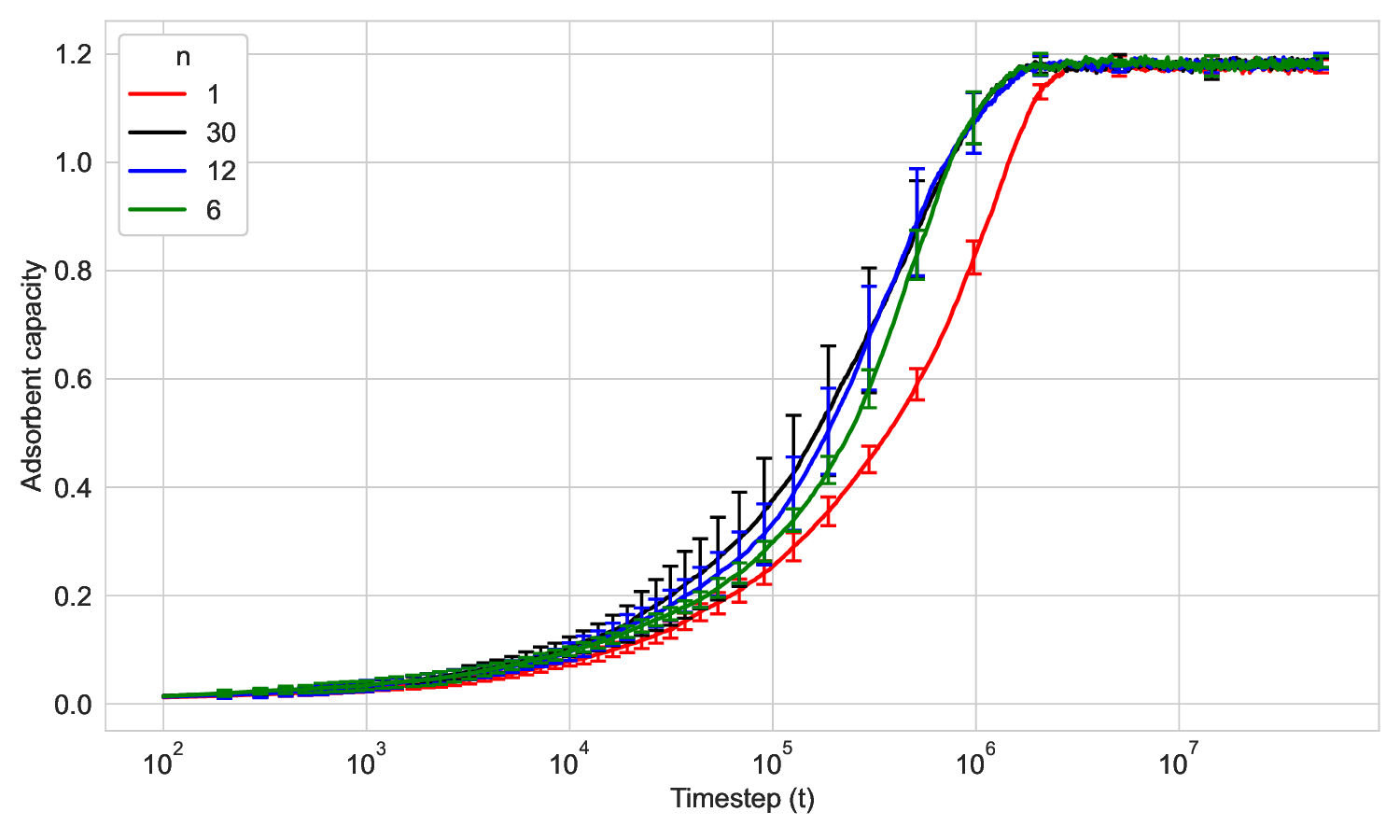}
\includegraphics[width=60mm]{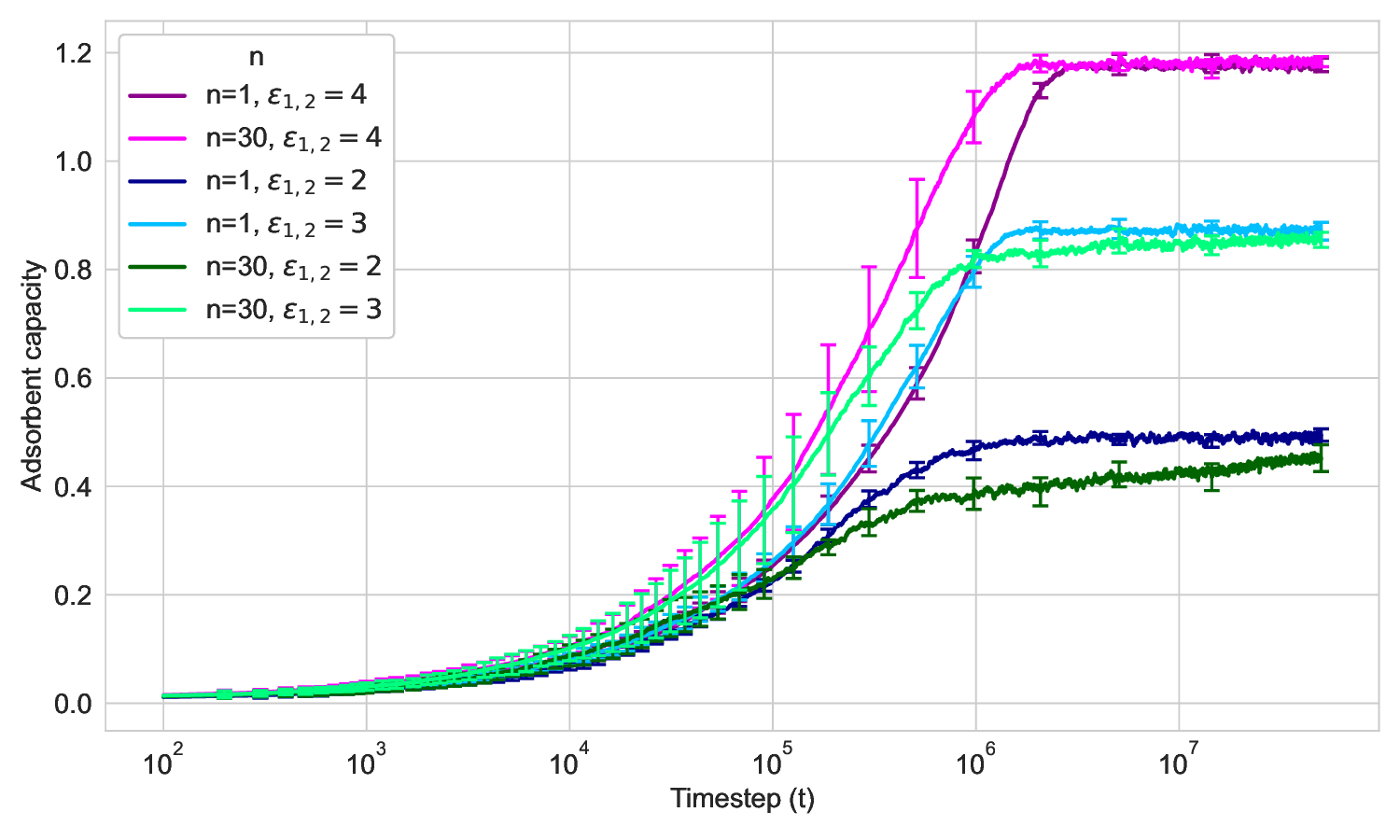}\\
(a)\hspace{14em}(b)\\
\caption{The time evolution of adsorption capacity, $q(t)$, in a log-linear scale for a set of polymer-particle mixtures.
(a) Comparison across the cases with different chain length $N$ (the number of chains is $n_p=600/N$ in each case) at fixed attraction strength $x_{ap} = 4$ in (\ref{ap}). (b) Comparison between two limiting cases, $n_p=1$ ($N=600$), and $n_p=30$ ($N=20$), only but for a range of well depths  $x_{ap}=2-4$. }
\label{eps=4}
\end{figure} 

The time evolution of adsorption capacity, $q(t)$, is compared across the cases with different chain lengths, $N$, at fixed attraction strength $x_{ap} = 4$ in Fig.~\ref{eps=4} (a). The high-molecular-weight polymer, $N=600$, is characterized by the slowest dynamics in reaching the stationary state when compared against the cases with an ensemble of shorter chains of the length $N=20-150$. We note that, according to the snapshots shown in Fig.~\ref{snap1}-\ref{snap3}, the longest chain merges into a single complex faster than a set of shorter chains. However, according to Fig. ~\ref{eps=4} (a), the $q(t)$ in this case stabilizes more slowly. This indicates that there is no direct relation between merging into a single complex and achieving stabilization of $q(t)$ for the longest chain. After a quick merge into a single complex, the growth of $q(t)$ still continues until saturated. The role of the attractive strength, given by $x_{ap}$, on the time evolution of $q(t)$ is shown in Fig.~\ref{eps=4} (b). Here two limiting cases, the longest, $N=600$, and the shortest, $N=20$ chains are shown only, and other cases fall in between these two. We make two remarks here. Firstly, the stationary state $q(\infty)$ depends solely on the well depth of the  polymer-particle interaction, $x_{ap}$, and its value increases non-linearly with $x_{ap}$: $q(\infty)\approx 0.5$, $0.85$, and $1.18$ for $x_{ap}=2$, $3$, and $4$, respectively. Secondly, the characteristic time for reaching a stationary state for the longest chain, $N=600$, is practically independent of the polymer-particle interaction strength, $x_{ap}=2-4$, whereas the shortest chains, $N=20$, reach this state essentially more slowly for the case of a weak attraction, $x_{ap}=2$.

\begin{figure}
\begin{center}
\includegraphics[width=60mm]{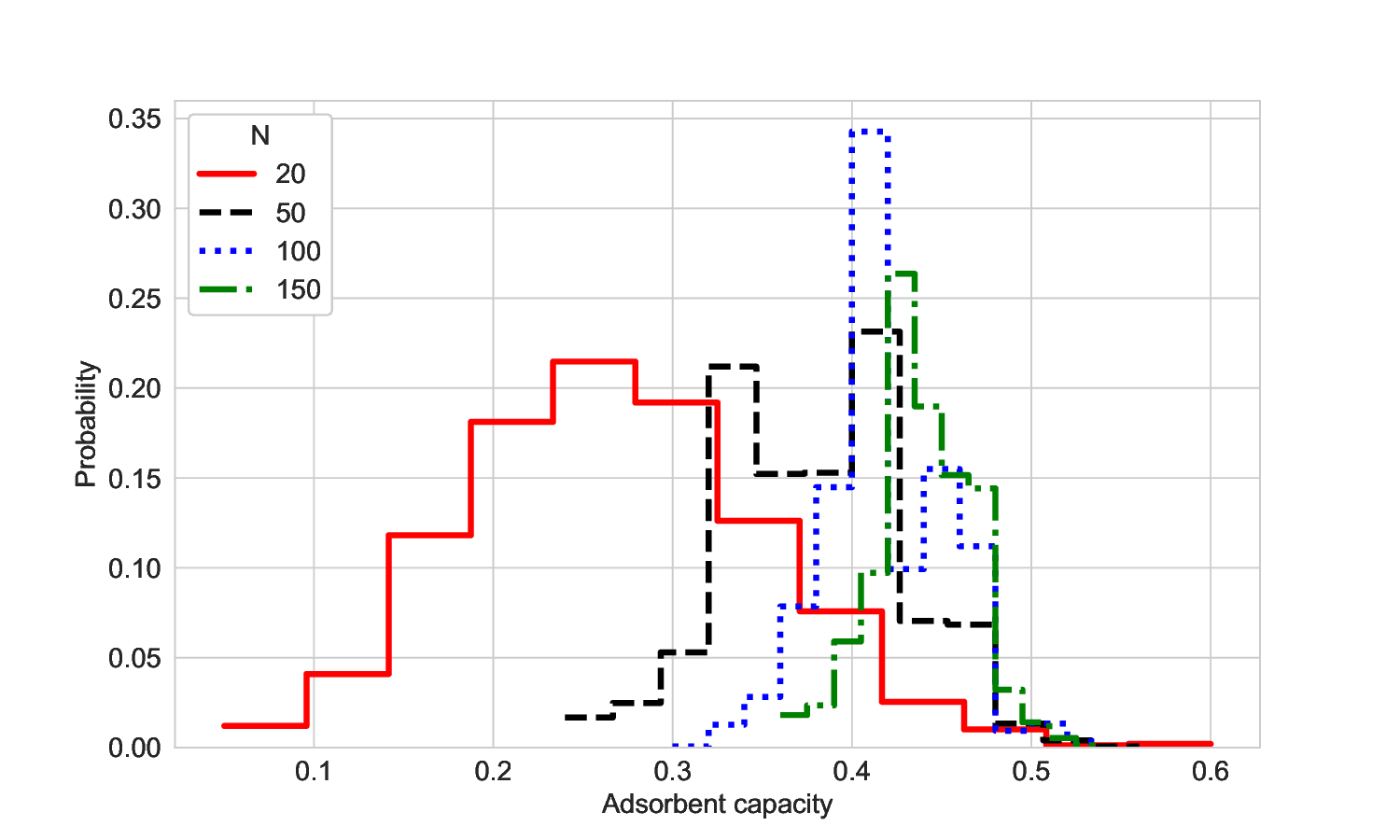}~
\includegraphics[width=60mm]{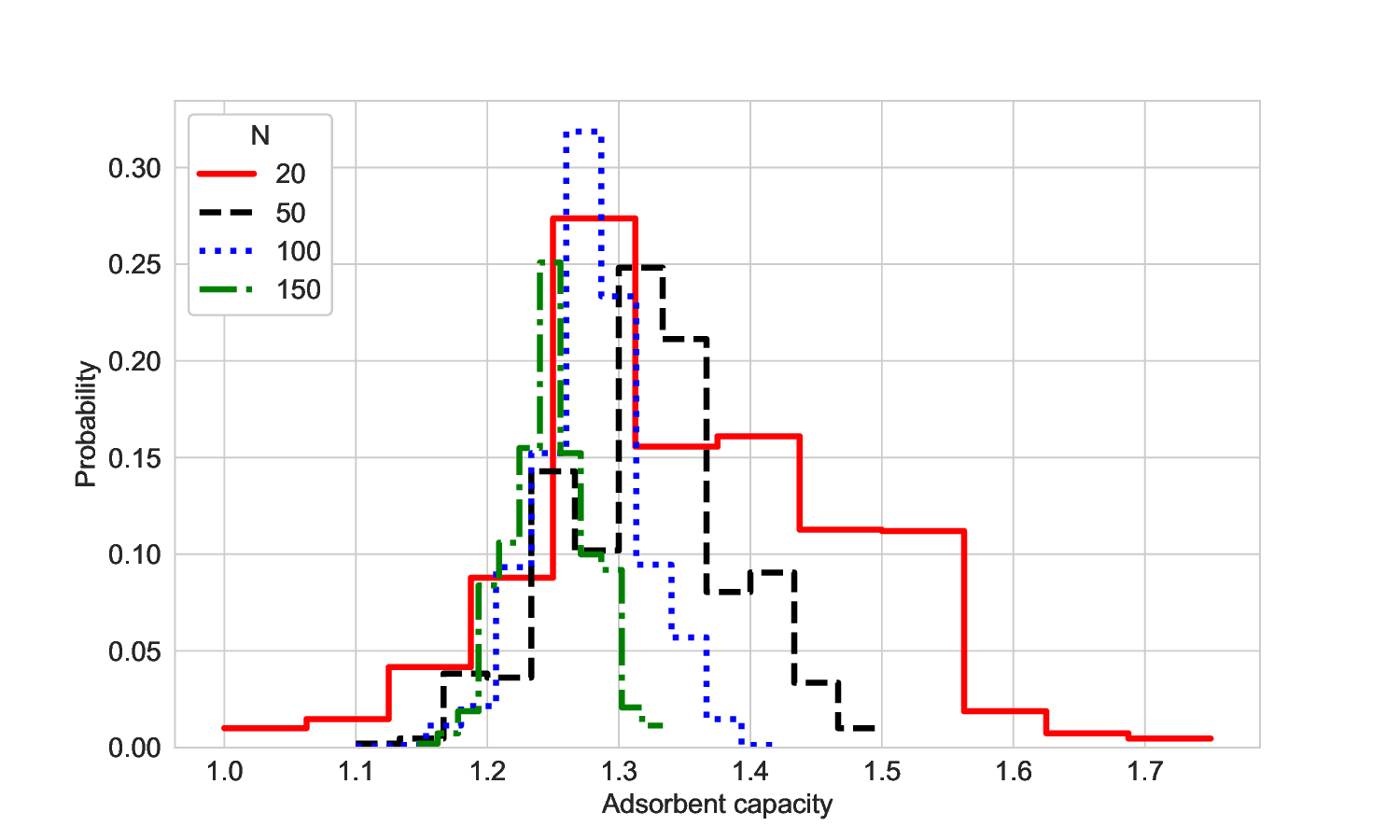}\\
(a)\hspace{14em}(b)\\
\end{center}
\caption{Probability distribution ($p(q)$) for adsorbent capacity in the stationary state for a single chain. (a) $x_{ap}=2$, (b) $x_{ap}=4$. }
\label{cap_dist_one_molecule}
\end{figure}

To gain more insight on these effects, we evaluated probability distribution, $p(q)$, for the $q(t)$ values in the stationary state for a single chain of variable length $N=20-150$, see Fig.~\ref{cap_dist_one_molecule}. A short chain of $N=20$ beads is characterized by a very broad probability distribution, indicating more frequent adsorption-desorption cycles. As the chain length increases, this distribution narrows down, and the average adsorption capacity increases, presumably because longer chains acquire enough conformational freedom to establish multiple contacts with an individual adsorbed particle. This holds for both $x_{ap}=2$ and $4$ cases; see frames (a) and (b). The position of the histogram maximum shifts essentially with the increase of a chain length for $x_{ap}=2$, but is practically unchanged for $x_{ap}=4$, in agreement with the conclusions drawn from Fig.~\ref{eps=4}.

\begin{figure}
\centering
\includegraphics[width=60mm]{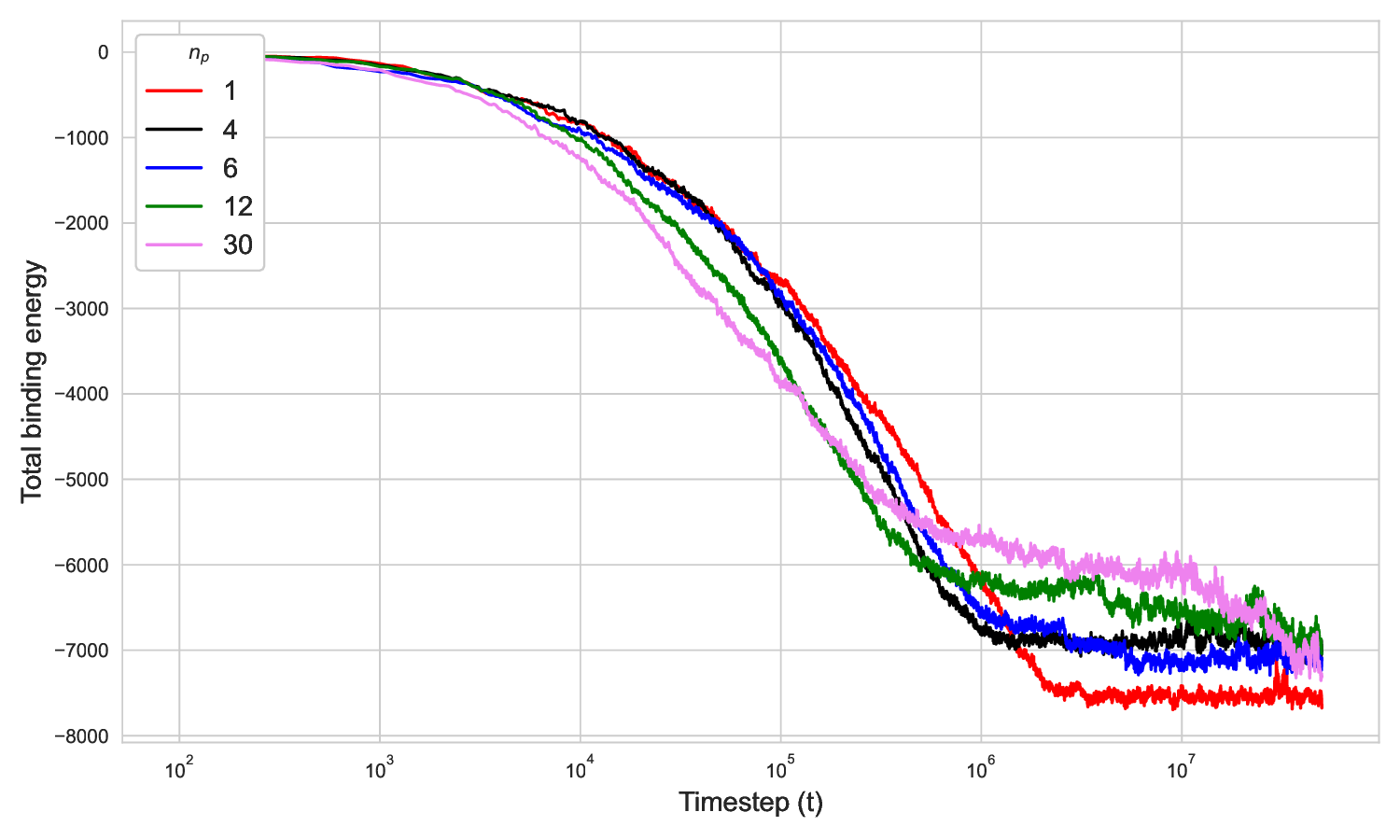}
\includegraphics[width=60mm]{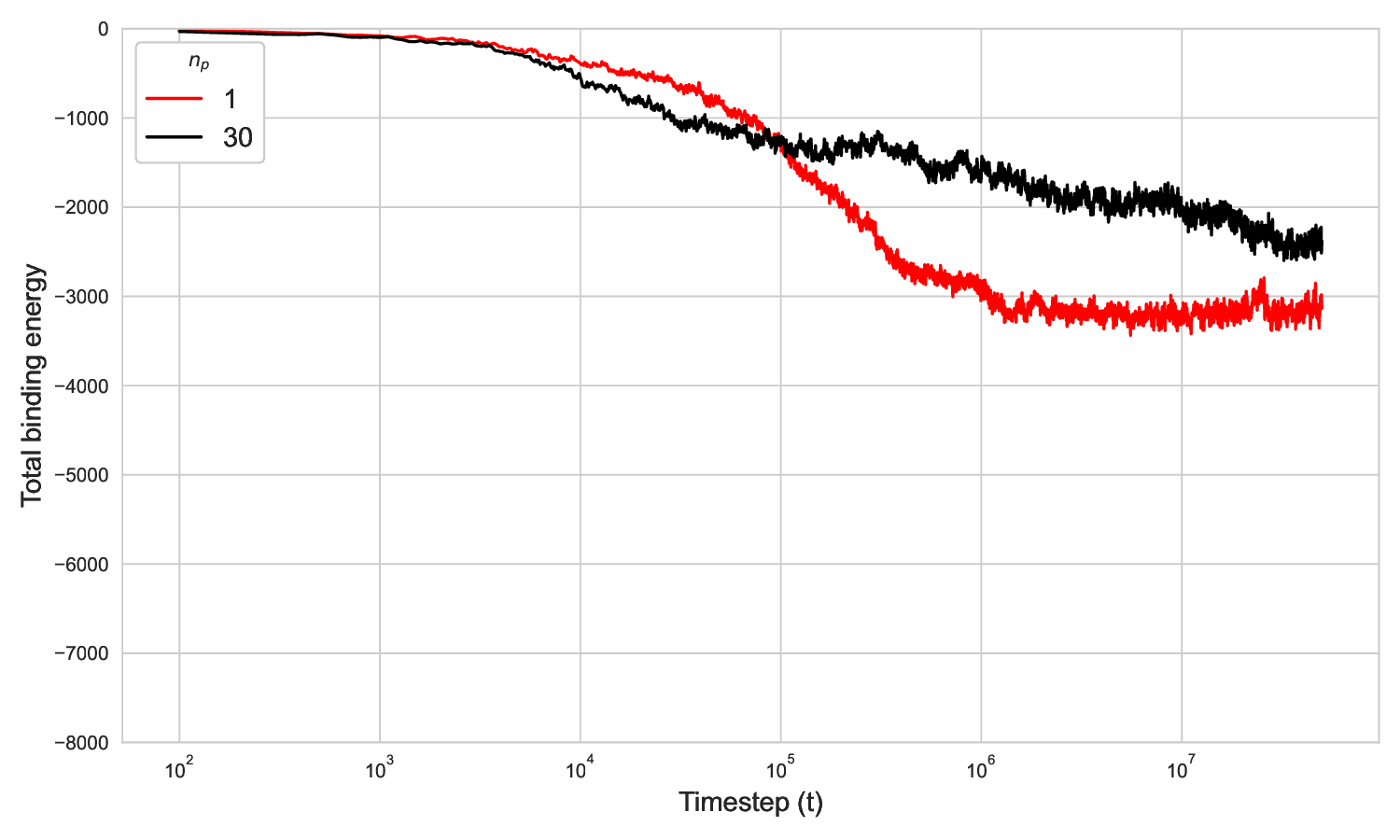}\\
(a)\hspace{14em}(b)\\
\caption{(a) Total binding energy  ($E_\mathrm{bind}$) for the polymers of various chain length, $N$, for $x_{ap}=4$. (b) The same for $x_{ap}=2$.}
    \label{energy}
\end{figure}

We examine the total binding energy $E_{{\rm bind}}$ of the polymer–particle complex next, see Fig.~\ref{energy}. While the adsorption capacity quantifies the number of particles that can be adsorbed under given conditions and increases with  time as shown above, the total binding energy represents the sum of all pairwise interactions between the polymer and the particles. We observe clearly two dynamics regimes here. During the first one, spanning from $t=0$ to about $5\cdot 10^5$, binding energy decreases quicker for short chains. We associate it with the preliminary formation of local complexes; see Fig.~\ref{snap1} and \ref{snap2}. It is faster for shorter chains as in this case it involves fewer conformation changes that cost an entropy penalty. During the second regime, spanning from $t=5\cdot 10^5$ to about $5\cdot 10^7$, local complexes merge into a single agglomerate. The merging mechanism for an ensemble of short chains is entirely diffusion-driven, and as the complexes gradually grow in size their self-diffusion slows down. In the case of a long chain, nuclei are necklaced along the chain promoting their faster merge; as a result, the total binding energy reaches a plateau faster. We note that for all the cases except the $N=600$ one, the total binding energy still does not reach a plateau within a time window of our simulations. 

\begin{figure}
\centering
\includegraphics[width=60mm]{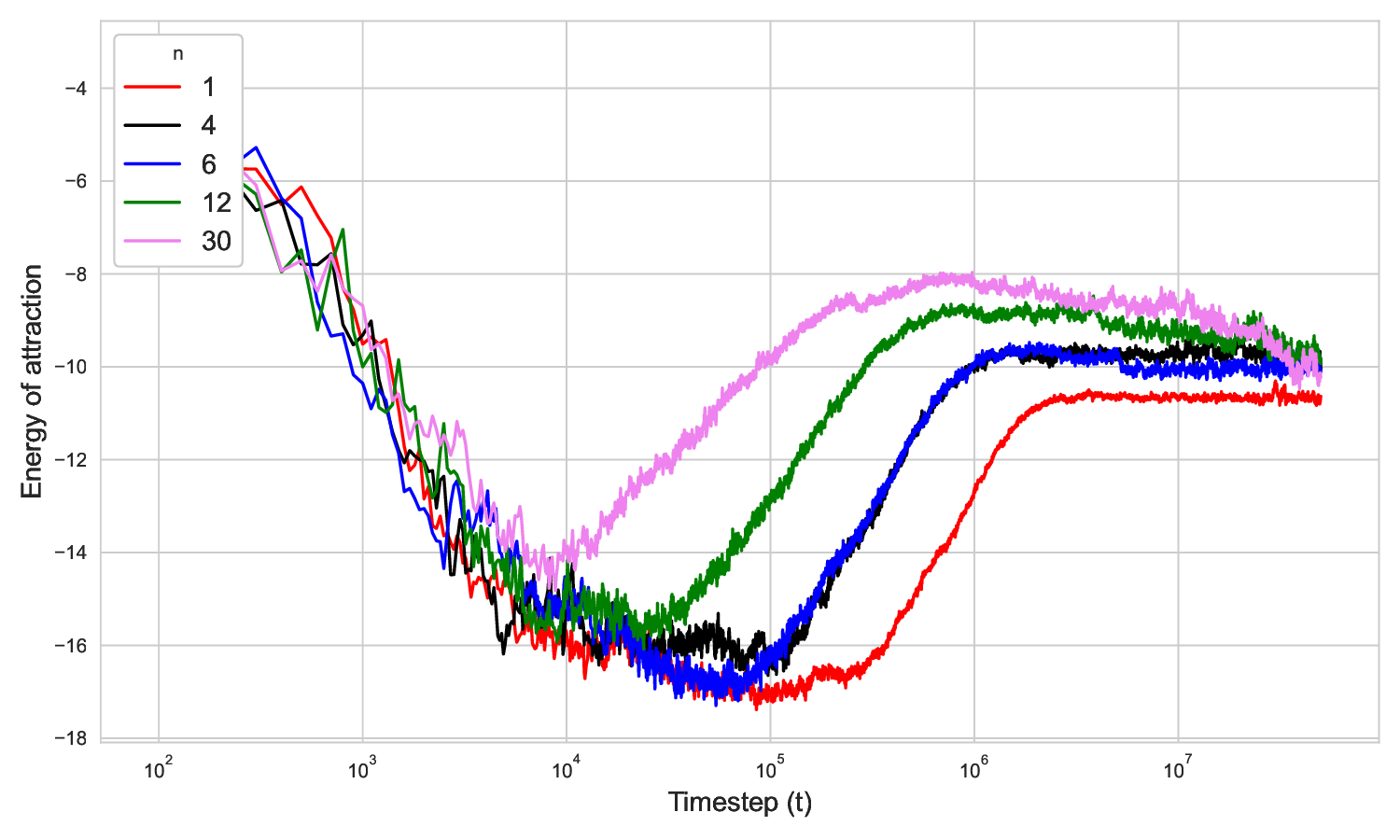} 
\includegraphics[width=60mm]{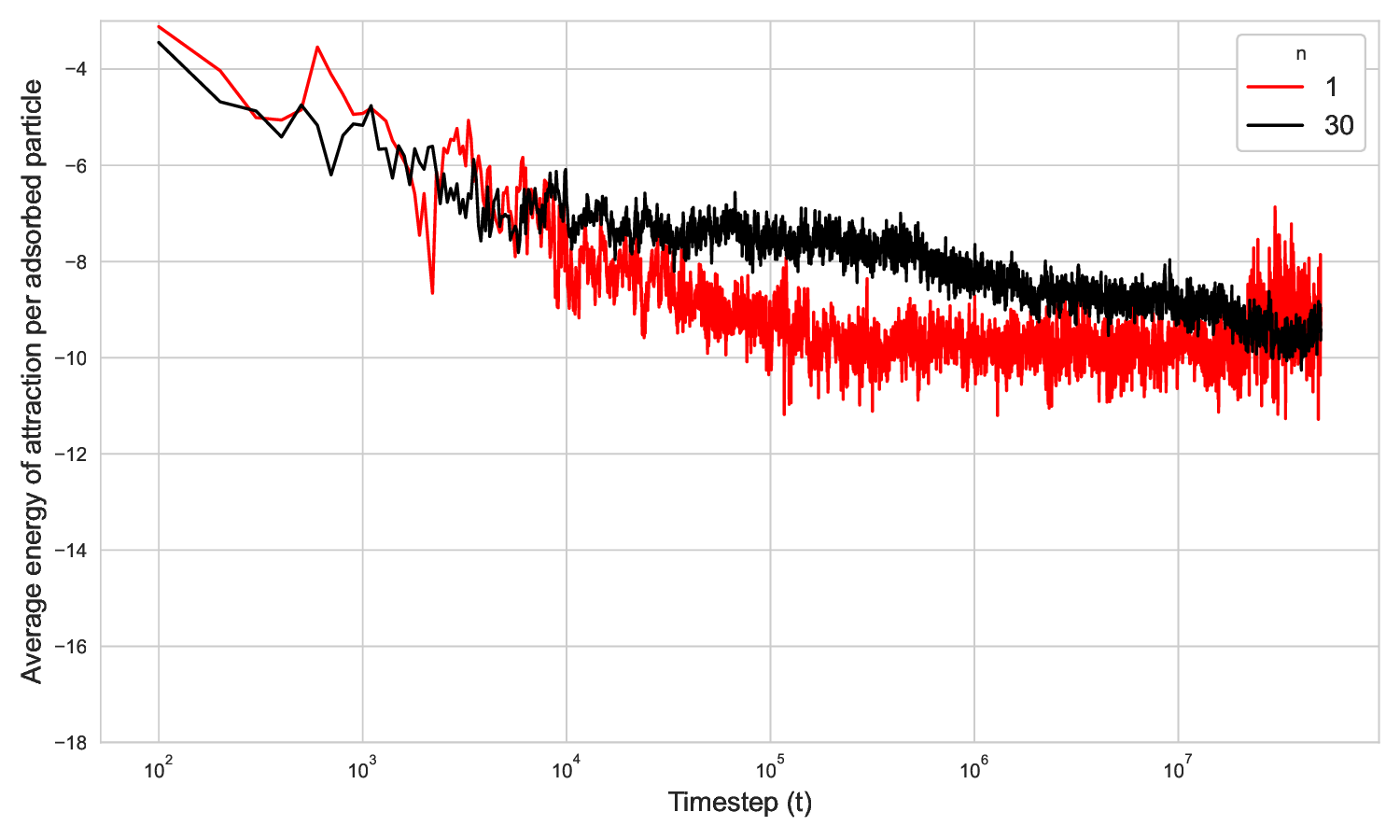}\\
(a)\hspace{14em}(b)\\
\caption{(a) Binding energy per adsorbed particle  ($E_\mathrm{bind}/(n_a(0)-n_a(t))$) for $x_{ap}=4$ for various polymer systems. (b) The same for $x_{ap}=2$.}
    \label{energy_mix}
\end{figure}

We continue our analysis by focusing on the binding energy per adsorbed particle ($E_{{\rm bind}}/(n_a(0)-n_a(t))$), see Fig.~\ref{energy_mix}, which quantifies an average binding strength of a particle. At $x_{ap}=4$, see Fig.~\ref{energy_mix} (a), it exhibits a non-monotonic behavior. The strongest binding (lowest binding energy) is observed at intermediate timescales, presumably due to the abundance of adsorbed particles with multiple coordination bonds. Longer chains present more opportunities in this respect because of their higher conformational flexibility. As the polymer system adsorbs more particles, the binding strength gradually decreases. This decrease is in part related to adsorption of particles onto the surface of the complex, where they form fewer coordination bonds, as the complex interior is already occupied. For a single long chain, this decrease stops upon reaching a plateau, coinciding with the attainment of maximum adsorption capacity. For the set of short chains, internal rearrangements within individual clusters also occur, optimizing the overall binding energy. This transition takes place on the same timescale as the saturation of the adsorption capacity. The curves eventually converge toward the same asymptotic value, defined by the plateau of the single high-molecular-weight chain. One should note that even though the single large-molecular-weight polymer chain reaches its maximum adsorption capacity later than the set of short chains, it attains its optimal energetic state at an earlier stage.

As it is discussed above, in the course of particle diffusion and further adsorption on the polymer flocculant, the polymer chain gradually forms a compact globule-like complex stabilized by multiple monomer-particle contacts. The process of compactification starts with the formation of localized nuclei along the polymer backbone until finally an almost spherical single complex is formed. The competition between the effect of structural (collapse) transition and adsorption kinetics is connected with the thermodynamic competition between energetic and entropic factors. For stronger polymer-particle interaction ($x_{ap}=4$), the energy gain from the monomer-particle binding overcomes the entropic swelling of the polymer coil. Thus, the rapid conformational collapse into a compact globule-like polymer-particle agglomerate occurs. During this rapid compactification process a significant fraction of the monomers becomes sterically screened within the core of this dense complex. This complicates the further diffusion of particles inside the complex, effectively slowing down the adsorption kinetics. Thus, the complex continues to adsorb particles at a reduced rate. For a weaker polymer-particle interaction ($x_{ap}=2$), the entropy-driven swelling dominates and prevents a rapid compaction of the polymer macromolecule, which in this case is characterized by a higher degree of conformational mobility and flexibility.

This effect can be further described by evaluation of the asphericity of a polymer-particle agglomerate. The universal asphericity parameter $ A$ was introduced \cite{Aronovitz1986, Rudnick1986} to distinguish between a completely rigid, rod-like polymer conformation (where it takes the maximum value of 1) and an ideally spherical shape (with a minimum asphericity value of 0), thus obeying the inequality $0 \leq  A  \leq 1$.
	This parameter is defined based on the eigenvalue spectrum of the gyration tensor, $\bf{Q}$, of the polymer-particle agglomerate structure consisting of $n_pN$ monomers and $n=n_a(0)-n_a(t)$ adsorbed particles, with the components defined as

 	\begin{equation}
 	Q_{ab}=\frac{1}{2(Nn_p+n)}\sum_{n,m=1}^{Nn_p+n}(\alpha_{n}-\alpha_{m})(\beta_{n}-\beta_{m}),
 	\end{equation}
 	where $\{\alpha,\beta\}=\{x,y,z\}$ are the spatial positions of polymer beads and pollutant particles comprising the agglomerate.
 	The radius of gyration $R_g$ is given by trace of  $\bf{Q}$,
 	whereas the asphericity $A$ is evaluated according to:
 	\begin{equation}
 	A=\frac{1}{6}\sum_{i=1}^3 \frac{(\lambda_i-{\overline{\lambda}})^2}{ {\overline{\lambda}^2}}, \label{asfer2}
 	\end{equation}
 	with ${\overline{\lambda}}\equiv {\rm Tr}\, {\bf{Q}}/d$ being
 	the mean arithmetic eigenvalue of  the gyration tensor eigenvalues. 
 
\begin{figure}
	\centering
	\includegraphics[width=24em]{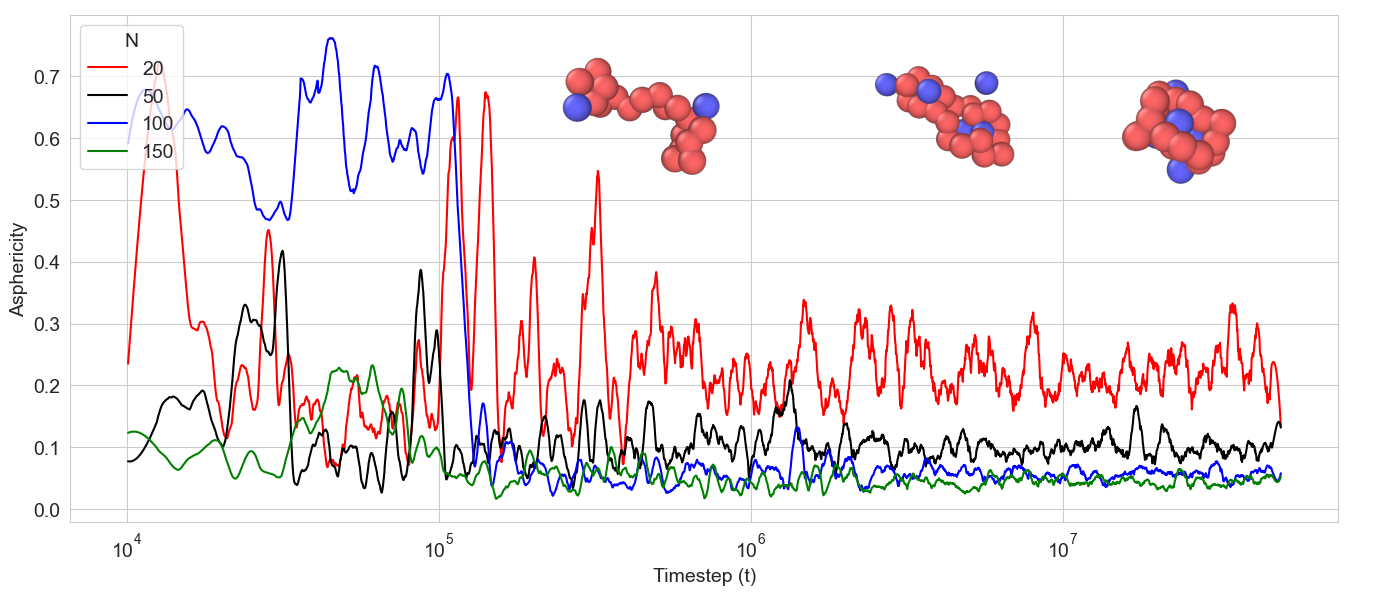}\\
	(a)\\
	\includegraphics[width=24em]{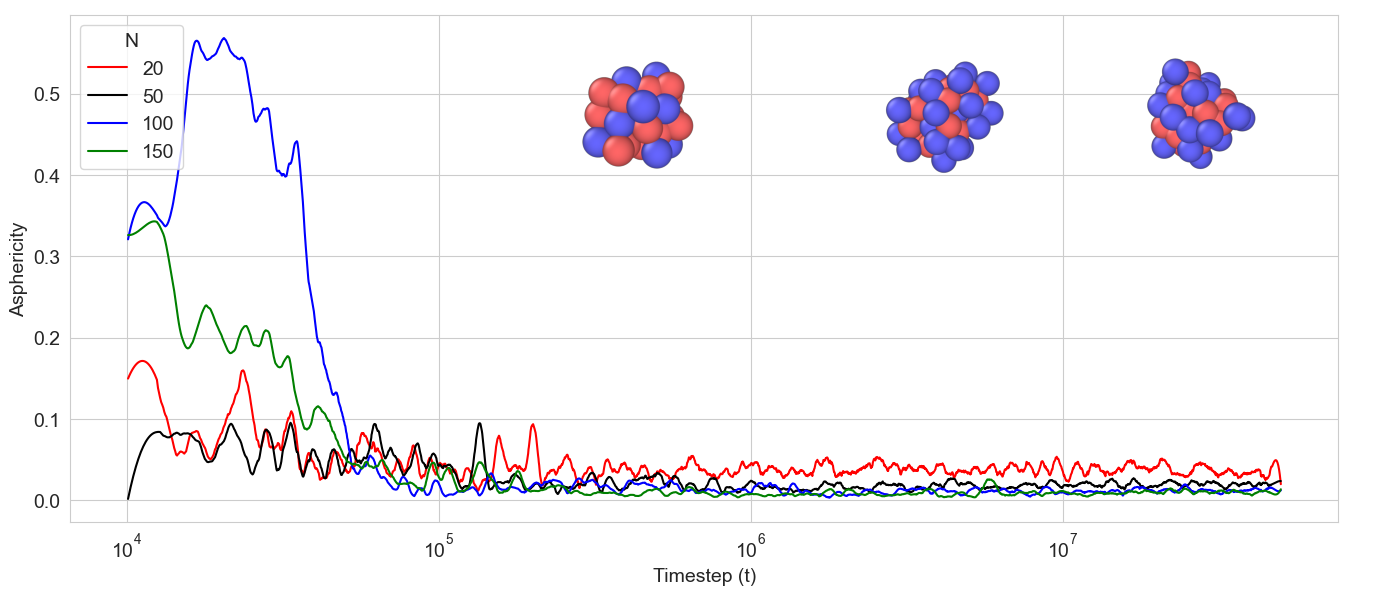}\\
	(b)\\
	\caption{(a) Asphericity of a single adsorbing chain for $x_{ap}=2$. (b) the same for $x_{ap}=4$. The curves are smoothed with the Savitzky-Golay filter for better readability. }
	\label{clust}
\end{figure} 

This quantity is strongly fluctuating for weak polymer-particle interaction, $x_{ap}=2$, indicating processes of collapse of polymer chains induced by adsorbed particles, followed by further swelling of polymers when some amount of adsorbed particles are released due to thermal fluctuations. As the chain length increases, the fluctuations are less pronounced, showing a much more stable formation of compact globule agglomerate; see Fig.~\ref{clust} (a). In the case of a strong polymer-particle interaction, $x_{ap}=4$, the difference between long and short chains is barely visible, as desorption processes are weaker under these conditions, and the bonds last longer, thereby supporting a long-term stable globule-like structure (see Fig. \ref{clust}b).

\begin{figure}
    \centering
    \includegraphics[width=60mm]{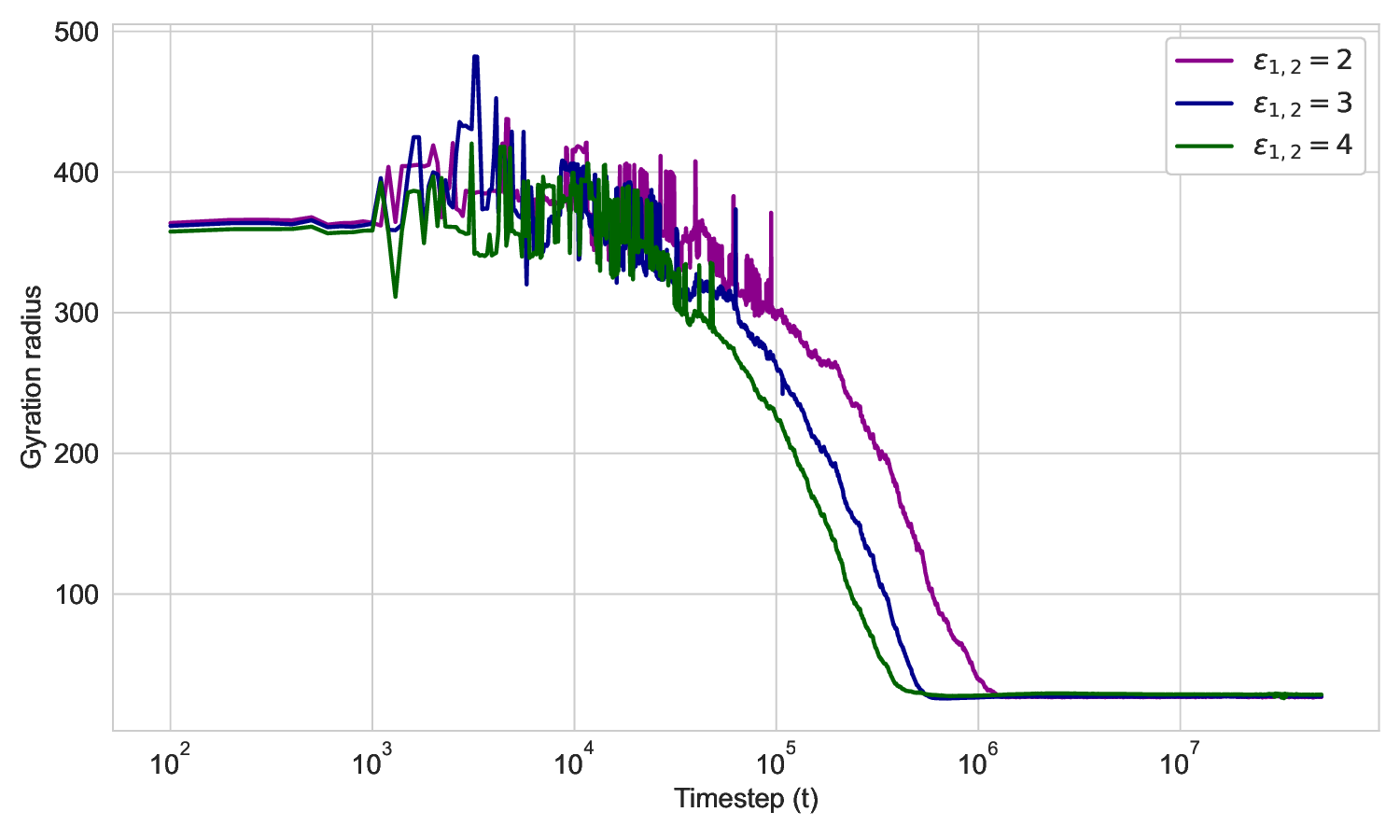}
    \includegraphics[width=60mm]{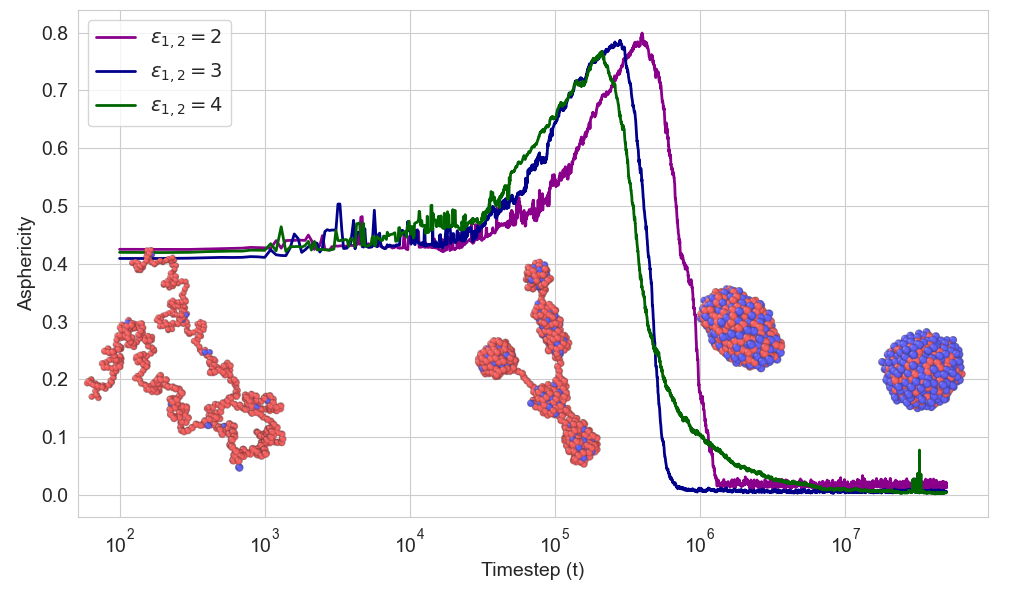}\\
(a)\hspace{14em}(b)\\
    \caption{(a) The time evolutions of the gyration radius {$R_g$} and (b) asphericity {$A$} in a log-lin scale for linear chain {of the length $N=600$} for different values of $x_{ap}$ averaged over 10 simulation runs. }
    \label{Chain}
\end{figure}
 
The presence of clusterization effects complicates the analysis of conformational properties; thus, the structural collapse of a set of polymer chains becomes much more difficult to study than in the case of the single linear chain. To investigate this collapse, we analyze the conformational size and shape properties, namely gyration radius $R_g$ and asphericity $ A$ of a single linear chain with $N=600$, as presented in Fig.~\ref{Chain}. The data presented in this figure are averaged over 10 simulation runs, each starting from a distinct random conformation.
The radius of gyration, $R_g$, gradually decreases with time, capturing the structural transition of the polymer chain to a compact state (Fig.~\ref{Chain}a). While strong attractive interactions accelerate this collapse, the maximum adsorption capacity is reached at a later stage (see Fig.~\ref{eps=4}). This indicates that the complex continues to adsorb pollutant particles even after the transition into a compact globular state is complete. The asphericity parameter provides a sensitive quantitative signature of the adsorption-induced collapse, exhibiting a distinct peak during the structural transition (Fig.~\ref{Chain}b). At the initial stage of adsorption, the asphericity attains a value slightly above $0.4$, which closely matches the known value for a swollen flexible coil in a good solvent regime: $A=0.431(2)$
\cite{jagodzinski1992}. During the second stage of  dynamics process, the small nuclei are being formed along the chain backbone, which leads to effective stretching of polymer conformations and thus an increase of asphericity. The parameter reaches its maximum value when only a few intermediate nuclei remain, where the high asphericity stems from the merging of these nuclei into a highly elongated conformation. Finally, a sharp drop in asphericity is observed when a single dense agglomerate is formed. Interestingly, stronger attraction slows down this final stage (as shown by the green curve). This occurs because the polymer-adsorbent attraction creates a higher kinetic barrier, making it more difficult for the system to attain the optimal spherical structure.

\begin{figure}
    \centering
    \includegraphics[width=120mm]{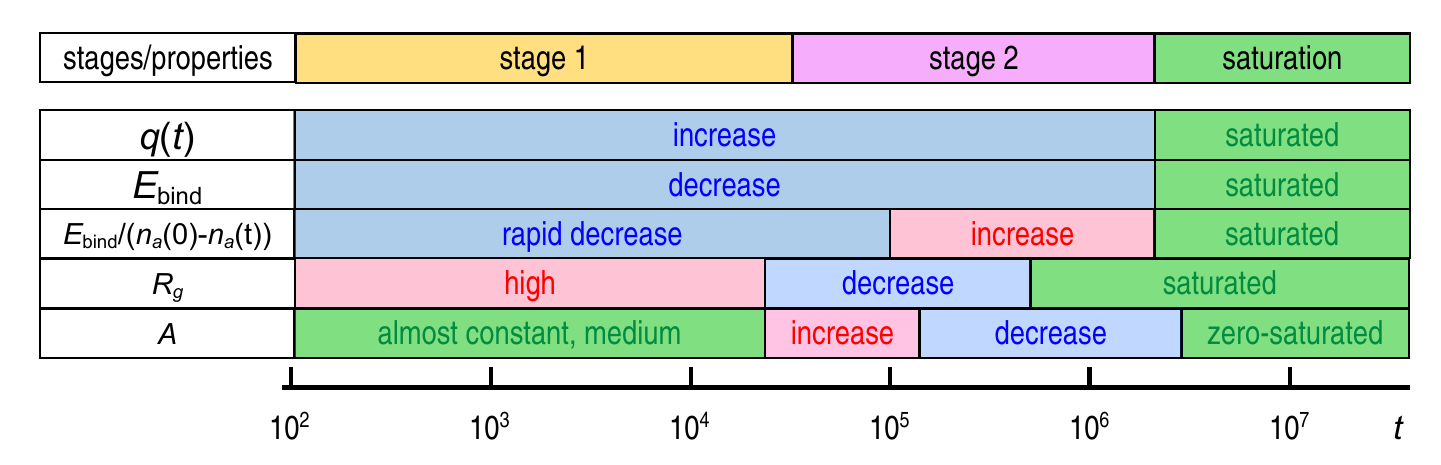}
    \caption{Schematic representation of the time-resolved adsorption stages for the case of a single chain of the length $N=600$ at the interaction potential well depth of $x_{ap}=4$.}
    \label{Schema}
\end{figure}
Our findings are systemized in a schematic form in Fig.~\ref{Schema} for the case of a single chain of the length $N=600$ at the interaction potential well depth of $x_{ap}=4$. One can distinguish three stages: stage 1, stage 2, and saturation with approximate boundaries as shown in this schema. For both adsorption capacity, $q(t)$, and total binding energy, $E_\mathrm{bind}$, stages 1 and 2 merge into a single one, where one observes both gradual monotonic increase of the number of adsorbed particles and respective decrease of total adsorption energy. These global characteristics, however, do not show some fine details that occur during this combined stage, and require analysis of the evolution of the energy per adsorbed particle, $E_\mathrm{bind}/(n_a(0)-n_a(t))$, gyration radius, $R_g$, and asphericity, $A$. In particular, at the stage 1 there is a rapid increase of $E_\mathrm{bind}/(n_a(0)-n_a(t))$ as the initially small number of adsorbed particles quickly form high coordination number nuclei with polymer beads. Both $R_g$ and $A$ at this stage remain almost constant, as only local conformational changes occur in the polymer not affecting its globally coiled state yet. At the stage 2, the nuclei along a chain gradually merge into larger {aggregates}. $E_\mathrm{bind}/(n_a(0)-n_a(t))$ increases, as the average coordination number within these aggregates decreases because of sharing the same polymer beads between multiple particles within the aggregate bulk. While growing in size, the aggregates are getting more compact in size as reflected by the decrease of $R_g$. Asphericity $A$ shows a maximum at the early times of the stage 2, reflecting abundance of temporarily highly aspherical objects, as the result of nuclei merging. Finally, at the later part of stage 2, $A$ decreases, reflecting formation of a final highly spherical aggregate. This scheme holds, to great extent, for other cases of $N$ and $x_{ap}$ with respective shifts of the stages boundary. These can be elucidated for the plots presented above.

\section{Conclusions}

We applied the Langevin dynamics to examine the kinetics of adsorption of low-molecular-weight particles (pollutants) onto flexible polymer macromolecules. The study can be considered as a natural continuation of our previous Monte Carlo simulations of the related problem of particle flocculation on polymer adsorbents, based on simple lattice models \cite{Blavatska2024a,Blavatska2024b}. The current study extends the analysis being reported in these previous works in two directions: (i) the use of an off-lattice model, and (ii) the examination of the adsorption dynamics depending on the molecular weight of the polymer adsorbent and of the attraction strength between the polymer and a pollutant.

Kinetics of adsorption can be split into three characteristic regimes: initial, intermediate, and saturation. During the former, small local polymer-particle (polymer-pollutant) complexes (nuclei) are formed, aided by the attraction between these two components. It is faster for short chains, as in this case the complexes form independently in bulk, and rather small conformational changes are required for foiling of such a chain around a particle. Long chains are characterized by the formation of a necklace-like structure of similar complexes (nuclei) arranged along the chain, and this process requires massive conformational rearrangements, and hence, it is slower. At the final adsorption stage, these local complexes merge into a single agglomerate. The situation in this case is reversed. Independent local complexes (in the case of short chains) merge solely due to their self-diffusion and random collisions between them, which slows down upon their gradual merge into larger complexes. A single necklace-like structure of a long chain with adsorbed pollutant merges quicker because of the topological connectivity of individual nuclei.

The structural transition of the polymer from a swollen coil to a compact pollutant-polymer agglomerate in the course of dynamic adsorption impacts its size and shape characteristics. The universal asphericity parameter $ A$ exhibits a non-monotonic behavior with a prominent peak during the intermediate stage of the process, where the highly elongated conformations appear temporarily before relaxing into a mostly spherical globular shape. Whereas a strong particle-polymer interaction regime ($x_{ap}=4$) accelerates the initial compaction, it introduces a higher kinetic barrier that slows down this final spherical rearrangement. We systemized these findings into a convenient time-resolved diagram, presented for the case of long chain, $N=600$, and $x_{ap}=4$, where all properties of interest are analyzed and put into the context of a whole picture of observed events.

From a practical perspective, our results are useful in selecting the optimal molecular weight of  polymeric adsorbents targeting specific classes of environmental pollutants (toxins). We found that for the rapid removal of highly hydrophobic or ionic pesticides (which correspond to regime of strong interactions), the set of relatively short polymer chains is preferable. On the other hand, for adsorbing the highly mobile, polar or weakly interacting organic pollutants, such as certain hydrophilic pesticides in  soil and groundwater, the long high-molecular-weight polymer chains are more effective. Their high conformational flexibility enables the formation of  binding pockets that structurally lock the mobile pollutant particles  and suppress the desorption fluctuations. 


\section*{Acknowledgements}
The work is supported by the Project No. 2025.07/0366 of the National Research Foundation of Ukraine " Functional efficiency of complex multibranched polymers in adsorption-based applications: computer simulations and predictions ".

\bibliographystyle{unsrt}
\bibliography{pom-pom_2}

\end{document}